\documentclass[aps,pra,reprint,showpacs]{revtex4-1}
\usepackage[utf8]{inputenc}
\usepackage{eurosym,amssymb,amsfonts,setspace,graphicx,bm,float,amssymb,tensor,amsmath,siunitx}
\newcommand{\pd}[1]{\frac{\partial}{\partial #1}}
\newcommand{\spd}[1]{\frac{\partial^2}{\partial #1^2}}

\newcommand{\expect}[1]{\langle #1 \rangle}

\newcommand{\dif}{\mathrm{d}}

\begin{document}
\title{Applications of the Fokker-Planck equation in circuit quantum electrodynamics}
\author{Matthew Elliott}
\affiliation{Advanced Technology Institute and Department of Physics, University of Surrey, Guildford GU2 7XH, United Kingdom}
\author{Eran Ginossar}
\affiliation{Advanced Technology Institute and Department of Physics, University of Surrey, Guildford GU2 7XH, United Kingdom}
\date{\today}
\begin{abstract}
 We study exact solutions of the steady state behaviour of several non-linear open quantum systems which can be applied to the field of circuit quantum electrodynamics. Using Fokker-Planck equations in the generalised $P$-representation we investigate the analytical solutions of two fundamental models. First, we solve for the steady-state response of a linear cavity that is coupled to an approximate transmon qubit and use this solution to study both the weak and strong driving regimes, using analytical expressions for the moments of both cavity and transmon fields, along with the Husimi $Q$-function for the transmon.  Second, we revisit exact solutions of quantum Duffing oscillator which is driven both coherently and parametrically while also experiencing decoherence by the loss of single and pairs of photons. We use this solution to discuss both stabilisation of Schr\"odinger cat states and the generation of squeezed states in parametric amplifiers, in addition to studying the $Q$-functions of the different phases of the quantum system. The field of superconducting circuits, with its strong nonlinearities and couplings, has provided access to new parameter regimes in which returning to these exact quantum optics methods can provide valuable insights.
\end{abstract} 
\maketitle

\section{Introduction}
The Fokker-Planck equation (FPE) is a valuable tool for finding exact steady-state solutions of driven, dissipative quantum oscillators. Most famously is has been used to treat the degenerate parametric amplifier \cite{Milburn1981,Drummond1981} and the quantum Duffing oscillator \cite{Drummond1980}. Such analytical solutions are particularly valuable to the study of quantum systems as they allow regimes to be studied where numerical simulation becomes unfeasible, for example very strongly driven systems where the Fock-state basis required for simulation become very large. They also enable large areas of parameter space to be studied very quickly. As experimental setups become more complicated, including multiple oscillators, there is increasing desire for solutions that help to study these systems. This becomes even more challenging when significant nonlinearities are also present in the system. Situations where steady state solutions of the FPE can be obtained, which is determined by whether the `potential conditions' are satisfied \cite{Walls2008}, are rare, making any new solutions that can be found of particular interest.

Superconducting quantum circuits \cite{Blais2004} give us the ability to conduct quantum optics experiments in a highly controlled and tunable environment where, unlike true atomic systems, we are free to design most of the parameters of the system. The Josephson junction provides strong nonlinearities enabling both the design of qubit circuits, such as the transmon \cite{Koch2007}, and efficient production of highly squeezed microwave fields \cite{Castellanos-Beltran2008}. The ability to create an effective 1D resonator which can be coupled almost perfectly to a transmission line also allows very efficient interaction between these squeezed fields and artificial atoms \cite{Murch2013}. Finally, the strong coupling that can be achieved between resonators and qubits gives us access to the strong dispersive regime \cite{Schuster2007,Gambetta2006a}, where the qubit can be used as a probe of the cavity state and vice versa, leading to the development of tomographic techniques in circuit quantum electrodynamics (circuit QED) \cite{Lutterbach1997,Bertet2002}. All these developments enable the study of parameter regimes which are inaccessible to conventional optics and it is therefore pertinent to revisit quantum optics methods to see how they may be adapted and extended to these new systems.

Current work in circuit-QED is particularly focused on scaling up to multi-oscillator systems, and optimal control is becoming increasingly relevant as devices improve in quality \cite{Puri2016,Tai2014}. In addition, there is great interest in using superconducting circuits to realise novel phases \cite{Boite2014,Jin2014} and quantum phase transitions \cite{Fitzpatrick2016} in driven dissipative lattices, while it is also hoped that a quantum simulator can be constructed from such an array \cite{Houck2012}. Efforts to improve the technology further have led to increased use of nonclassical states, for example for improving qubit read out \cite{Barzanjeh2014,Khezri2016}. The field of quantum optomechanics \cite{Bowen2016,Aspelmeyer2014} is concerned with the same fundamental models as circuit QED, albeit in different parameter ranges and can therefore also benefit from the methods discussed here. Much current work is focused on cooling a mechanical resonator into its ground state \cite{Yuan2015,Peterson2016,Habibi2016}, and the related problem of engineering a macroscopic vibrational superposition state \cite{Abdi2016}. Work is also being done on using a mechanical oscillator to more precisely characterise an optical mode \cite{Clark2016}, in addition to using the cavity to perform sensitive mechanical measurements \cite{Aasi2013,Pooser2015}. Cavity optomechanics also provides a novel method of converting between microwave and optical photons \cite{Andrews2014}, opening up the possibility for hybrid quantum information systems.

In this paper, we extend the FPE method to treat two systems of interest in circuit QED. First we study a transmon qubit, modelled as a quantum Duffing oscillator, coupled to a linear readout cavity. Using an adiabatic elimination process, we derive expressions for the steady-state moments of both the transmon and cavity fields, in addition to $Q$-functions of the transmon. We show that that despite the apparent restrictiveness of this process, we retain much of the important behaviour of the system in our effective single oscillator system, even when the cavity and qubit are resonant and this approach would seem most likely to break down. The Jaynes-Cummings model, which approximates the transmon as a two-level system, has been studied extensively using numerical solutions at low occupation and semiclassical models in the limit of strong driving \cite{Bishop2010} and in the presence of non-zero temperature \cite{Rau2004,Fink2010}. In the case of strong driving, however, the higher levels of the transmon become relevant to the dynamics and no analytical solution exists in this regime. The high power regime is of particular interest for performing high fidelity, fast qubit readout \cite{Reed2010}. We plot the analytic cavity and transmon response, in both the dispersive \cite{Schuster2007} and resonant \cite{Bishop2008} regimes, over several orders of magnitude of drive power, observing many features of the system that are seen experimentally.

Second, we consider a Duffing oscillator which is driven both coherently and parametrically, while decoherence occurs through the loss of both single and pairs of photons. This system, particularly the parametrically driven Duffing system, has been studied extensively and exact solutions for the moments of the field already exist, but returning to these models in the circuit-QED regime can provide new insights. For example, this model is important in the study of the period doubling bifurcation \cite{Dykman1998,Zorin2011} and is relevant to a proposed scheme for high-fidelity qubit readout \cite{Krantz2016}. We derive analytical expressions for the resonator $Q$-function to study the difference between the classical and fully quantum steady states of the parametrically driven system. In addition, we study that application of this model to a recent proposal to stabilise Schr\"odinger cat states in circuit QED \cite{Leghtas2015}, where we see that the distortions due to the cavity self-Kerr \cite{Kirchmair2013} induced by coupling to a qubit are significantly reduced by introducing a two-photon loss process. We also study how the presence of a quartic nonlinearity in an otherwise ideal parametric amplifier \cite{Milburn1981} affects the ability to generate intracavity squeezed states.

\section{The Cavity-Transmon system}
Superconducting qubits are nonlinear resonators which have sufficient large anharmonicity that the transition between the lowest two levels can be addressed selectively \cite{Clarke2008}. One such device is the transmon, which has greatly reduced charge noise compared with other qubits \cite{Koch2007} and can achieve long coherence times \cite{Paik2011} and is therefore widely used in experiments \cite{Wang2016,Suri2015,Riste2015}. Its relatively weak negative anharmonicity, when compared with atomic systems, however, means that at high drive powers additional levels beyond the computational basis must be considered, with the quantum Duffing oscillator providing a good approximation to the level structure \cite{Bishop2010a}. When coupled to a linear read-out cavity, the Hamiltonian for the full system is 
\begin{multline}
H_1 = \omega_c a^\dag a + i (\epsilon e^{-i \omega_d t} a^\dag  - \epsilon^* e^{i \omega_d t} a ) + ig(a b^\dag - a^\dag b)
\\
+ \omega_t b^\dag b + \frac\chi2 b^\dag b^\dag b b,
\end{multline}
in the rotating wave approximation, where $a$ and $b$ are the annihilation operators for the cavity and transmon modes which have frequencies $\omega_c$ and $\omega_t$ respectively, $\epsilon$ is the coherent drive strength, $\omega_d$ is the driving frequency, $g$ is the cavity-transmon coupling and $\chi$ is the transmon anharmonicity. In order to remove the time-dependence of the Hamiltonian we transform into a rotating frame at the drive frequency,
\begin{multline}
\tilde{H}_1 = \Delta_c a^\dag a + i (\epsilon a^\dag - \epsilon^* a) + ig(a b^\dag - a^\dag b)
\\
+ \Delta_t b^\dag b + \frac\chi2 b^\dag b^\dag b b,
\end{multline}
where we have defined $\Delta_c=\omega_c - \omega_d$ and $\Delta_t=\omega_t - \omega_d$. A master equation allows us to study the dynamics of this system under the influence of dissipation into a zero-temperature bath via both the cavity and transmon. This is given by

\begin{equation}
\dot{\rho} = -i[\tilde{H}_1,\rho] + \mathcal{L}[\sqrt{\gamma_c} a]\rho + \mathcal{L}[\sqrt{\gamma_t} b]\rho,
\end{equation}
where $\mathcal{L}[a] = a \rho a^\dag - \frac12 a^\dag a \rho - \frac12 \rho a^\dag a$ and $\gamma_c$ and $\gamma_t$ are the cavity and transmon decay rates respectively. We are interested in exact steady state solutions of this system and therefore rewrite this equation in the form of a FPE in the generalised $P$-representation \cite{Drummond1980a}, as has been used to solve other nonlinear cavity systems \cite{Walls2008}

\begin{multline}
\label{eqn:fpe}
\frac{\partial P_1(\pmb\alpha)}{\partial t} = \biggl[-\pd{\alpha_1} \left(-i\Delta_c \alpha_1 + \epsilon -g\alpha_2 - \frac{\gamma_c}2 \alpha_1 \right) 
\\
-\pd{\beta_1} \left(i\Delta_c \beta_1 + \epsilon^* - g\beta_2 - \frac{\gamma_c}2 \beta_1 \right)
\\
-\pd{\alpha_2}\left(g \alpha_1 - i \Delta_t \alpha_2 - i \chi \alpha_2^2 \beta_2 -\frac{\gamma_t}2 \alpha_2 \right)
\\
-\pd{\beta_2}\left(g \beta_1 + i \Delta_t \beta_2 + i \chi \beta_2^2 \alpha_2 -\frac{\gamma_t}2 \beta_2 \right)
\\
+\frac12 \spd{\alpha_2} \left(-i\chi \alpha_2^2\right) +\frac12 \spd{\beta_2}\left(i\chi \beta_2^2 \right) \biggr] P_1(\pmb\alpha),
\end{multline}
where $(\alpha_1$, $\beta_1)$ are the phase-space coordinates of the cavity, $(\alpha_2$, $\beta_2)$ are those of the transmon and $P_1(\pmb{\alpha})$ is a quasiprobability distribution over the phase space with $\pmb{\alpha}=(\alpha_1,\beta_1, \alpha_2, \beta_2)$. In the generalised $P$-representation, the $\alpha_i$ and $\beta_i$ need only be complex conjugate on average \cite{Drummond1980a}, and any moments much be found by integrating over the full 8-dimensional space.

\subsection{Adiabatic elimination of the cavity}
In this form of Eq. (\ref{eqn:fpe}) the steady state of the system cannot be solved for analytically by the potential conditions method. If  $\gamma_c \gg \gamma_t$, however,  then we can perform an adiabatic elimination of the cavity. We assume that the cavity is so fast that it relaxes instantaneously in response to changes in the transmon field and therefore remains in a steady state. Via a conversion to the form of a Langevin equation and back again, in a similar fashion to that used in \cite{Drummond1981}, we obtain relations for the coordinates of the cavity in terms of those of the transmon

\begin{figure}
    \centering
    \includegraphics[width=\columnwidth]{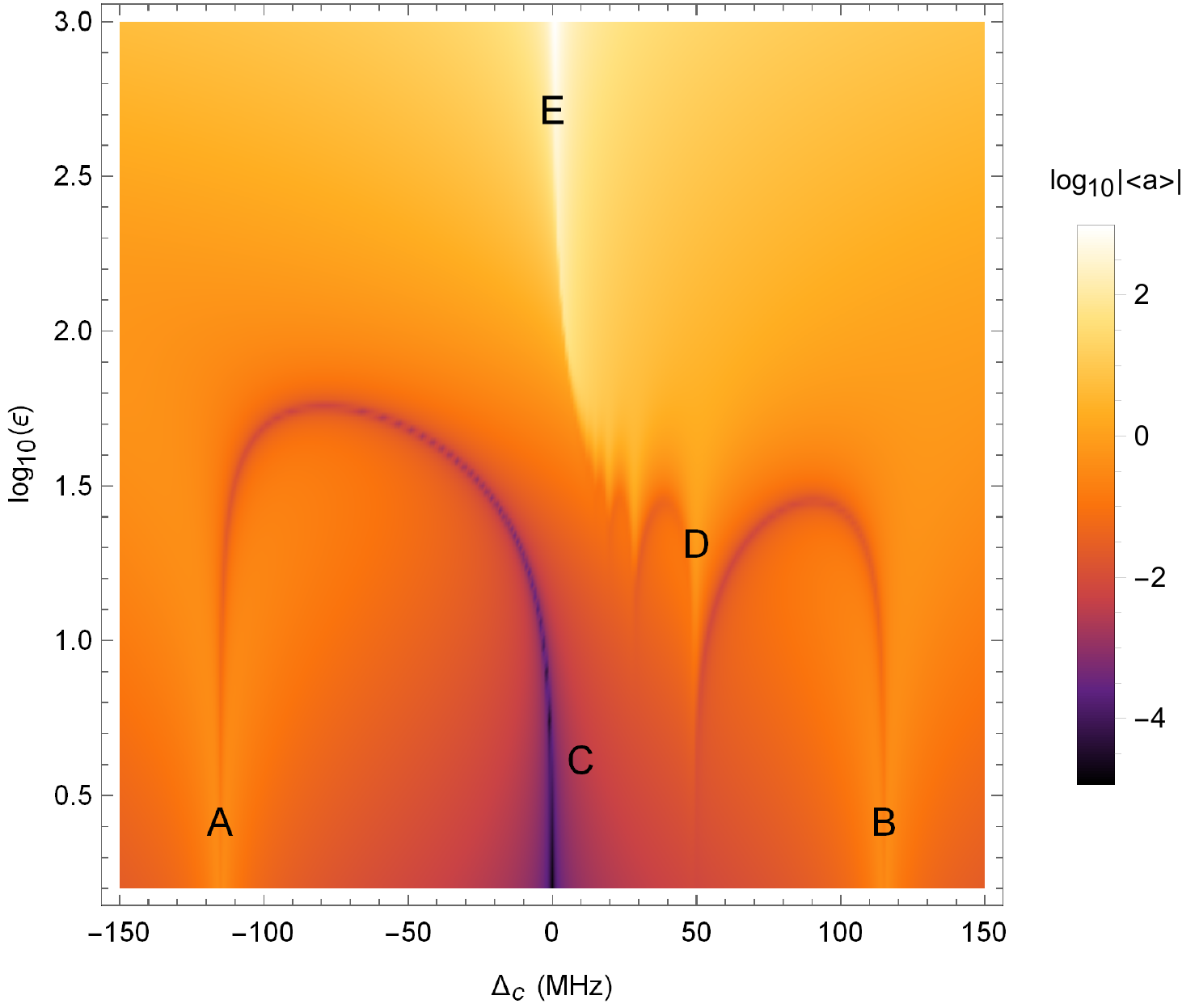}
    \caption{(Colour Online) Plot of $|\expect{a}|$ as a function of the detuning $\Delta_c$ of the drive from the bare cavity and drive amplitude $\epsilon$ when cavity and qubit are resonant. Other parameters are $g/2\pi=\SI{115}{\mega\hertz}$, $\chi/2\pi=\SI{-220}{\mega\hertz}$, $\gamma_c/2\pi=\SI{2}{\mega\hertz}$, $\gamma_t/2\pi=\SI{0.1}{\mega\hertz}$. We see the characteristic vacuum Rabi splitting, with peaks (A,B) separated by $2g$ at low powers and then demonstrating `supersplitting' as the power increases. There is extremely low transmitted amplitude at the bare cavity frequency (C). As the power increases, higher order transitions become present in the spectrum (D) and at sufficiently large drive strengths the resonance shifts back to the bare cavity frequency and there is a strong transmission peak (E).}
    \label{fig:resonant}
\end{figure}

\begin{figure}
    \centering
    \includegraphics[width=\columnwidth]{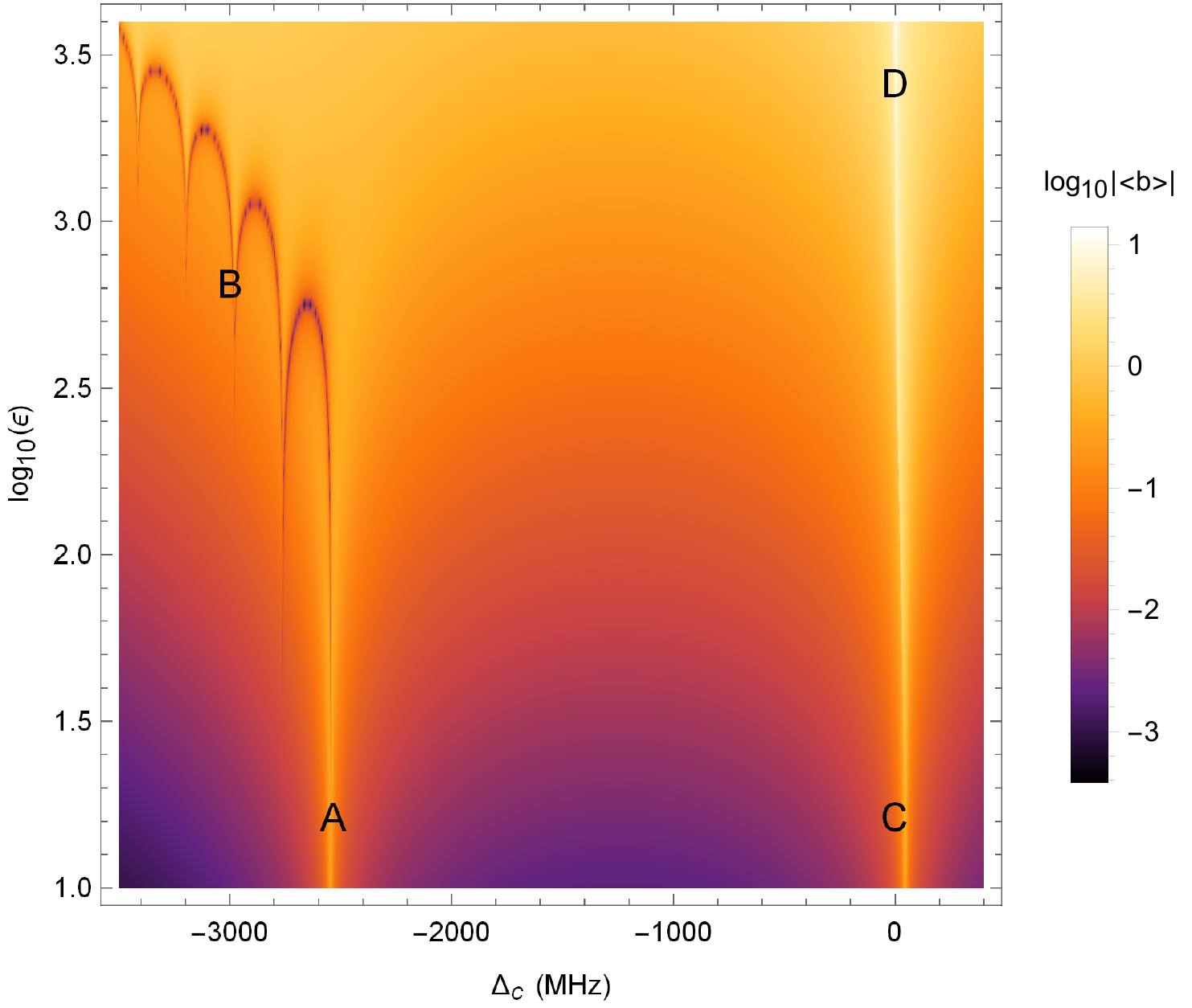}
    \caption{(Colour Online) Plot of $|\expect{b}|$ as a function of the detuning $\Delta_c$ of the drive from the bare cavity frequency and drive amplitude $\epsilon$ when cavity and qubit are coupled in the strong dispersive regime. Other parameters are $\Delta_{ct}=\SI{2.5}{\giga\hertz}$, $g/2\pi=\SI{340}{\mega\hertz}$, $\chi/2\pi=\SI{-220}{\mega\hertz}$, $\gamma_c/2\pi=\SI{2}{\mega\hertz}$ and $\gamma_t/2\pi=\SI{0.1}{\mega\hertz}$. Around the bare transmon frequency the system behaves like a quantum Duffing oscillator with a dispersively shifted fundamental frequency (A). At higher powers we see peaks corresponding to transitions between higher transmon levels which are separated by $\chi$ (B). Near the bare cavity resonance the transmission peak is dispersively shifted at low power (C), but shifts to the bare cavity frequency at high power (D). This region in shown in greater detail in Fig. \ref{fig:dispersive2}.}
    \label{fig:dispersive1}
\end{figure}

\begin{equation}
\alpha_1 = \frac2{\tilde{\gamma}_c}(\epsilon - g \alpha_2) ~~~~ \beta_1 = \frac2{\tilde{\gamma}_c^*}(\epsilon^* - g \beta_2),\label{eqn:elim}
\end{equation}
where we have defined $\tilde{\gamma}_c= \gamma_c + 2i\Delta_c$ (see Appendix \ref{app:elim} for full details). We substitute these relations back into the FPE to give the single-oscillator equation
\begin{multline}
\label{eqn:FPE1}
\frac{\partial P_1(\pmb\alpha)}{\partial t} = \biggl[-\pd{\alpha_2}\left(\tilde{\epsilon} - i \chi \alpha_2^2 \beta_2 -\frac{\tilde{\gamma}_t}2 \alpha_2 \right) 
\\
-\pd{\beta_2}\left(\tilde{\epsilon}^* + i \chi \beta_2^2 \alpha_2 -\frac{\tilde{\gamma}_t^*}2 \beta_2 \right)
\\
+\frac12 \spd{\alpha_2} \left(-i\chi \alpha_2^2\right) +\frac12 \spd{\beta_2}\left(i\chi \beta_2^2 \right) \biggr] P_1(\pmb\alpha),
\end{multline}
where we have additionally defined an effective decay constant for the transmon $\tilde{\gamma}_t = \gamma_t +2i\Delta_t  +2g^2/\tilde{\gamma}_c $ and an effective drive strength $\tilde{\epsilon} = 2g\epsilon/\tilde{\gamma}_c $. This is essentially the FPE for a driven, damped quantum Duffing oscillator \cite{Drummond1980} but with parameters which are inherently complex numbers. This simplified system does satisfy the potential conditions, which allows us to find an expression for the steady state moments of the transmon (further details in Appendix \ref{app:moments}),

\begin{multline}
\label{eqn:moments}
\expect{b^{\dag n}b^m } =
\\
\frac{\left(\frac{\tilde{\epsilon}}{i\chi}\right)^m\left(\frac{\tilde{\epsilon}^*}{-i\chi}\right)^n\Gamma(d)\Gamma(d^*) ~_0F_2(d+m,d^*+n,2|\frac{\tilde{\epsilon}}{\chi}|^2)}{\Gamma(d+m)\Gamma(d^*+n) ~_0F_2(d,d^*,2|\frac{\tilde{\epsilon}}{\chi}|^2)},
\end{multline}
where $\Gamma(x)$ is the Gamma function, $~_0F_2$ is a generalised hypergeometric function and we have defined $d=\tilde{\gamma}_t/2i\chi$. In addition it is possible to produce similar analytic expressions for the Fock state distribution $P(n)$ and the Husimi $Q$-function for the transmon mode, which are given in Appendix \ref{app:pandq}.

\begin{figure}
    \centering
    \includegraphics[width=.98\columnwidth]{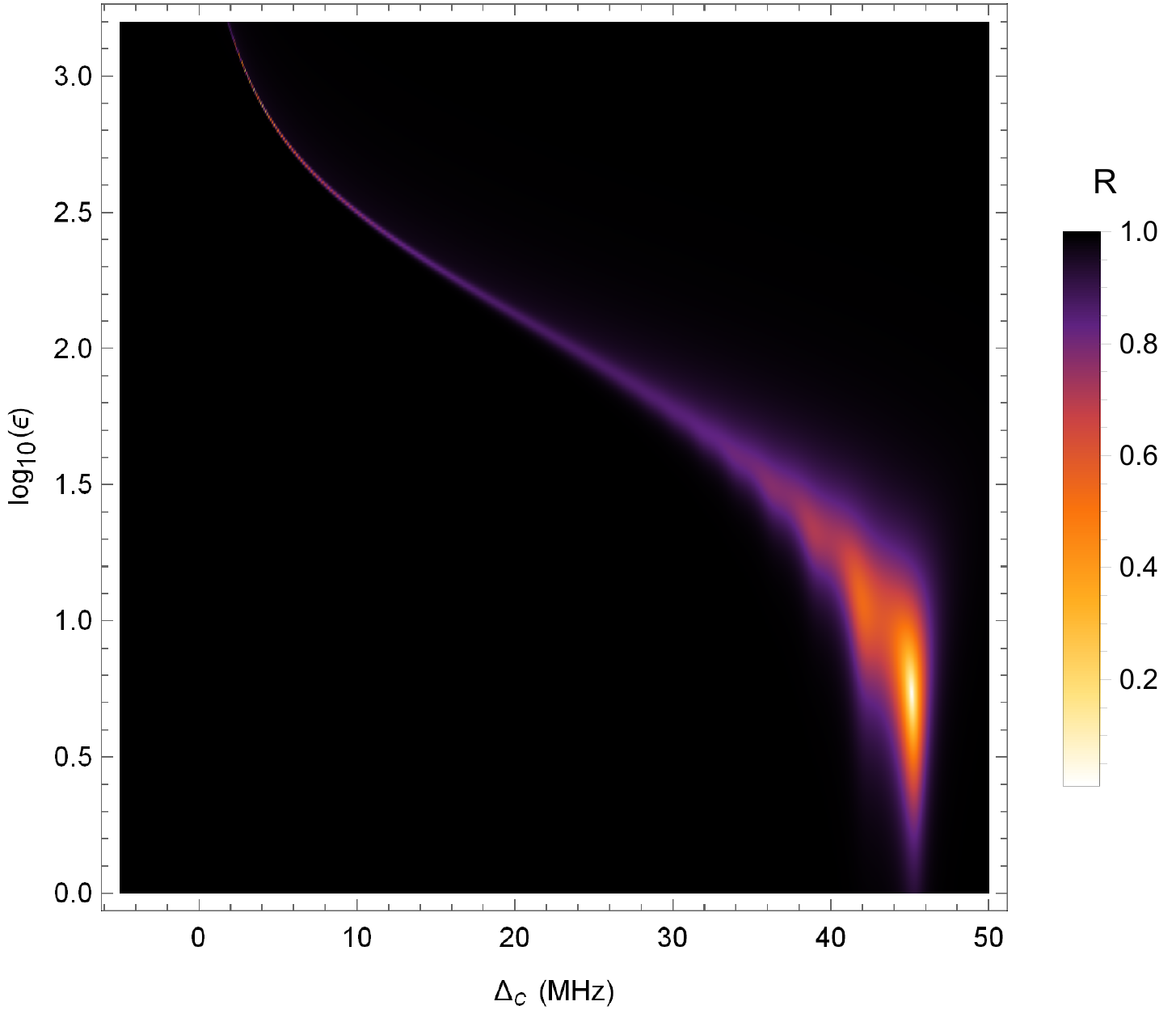}
    \includegraphics[width=\columnwidth]{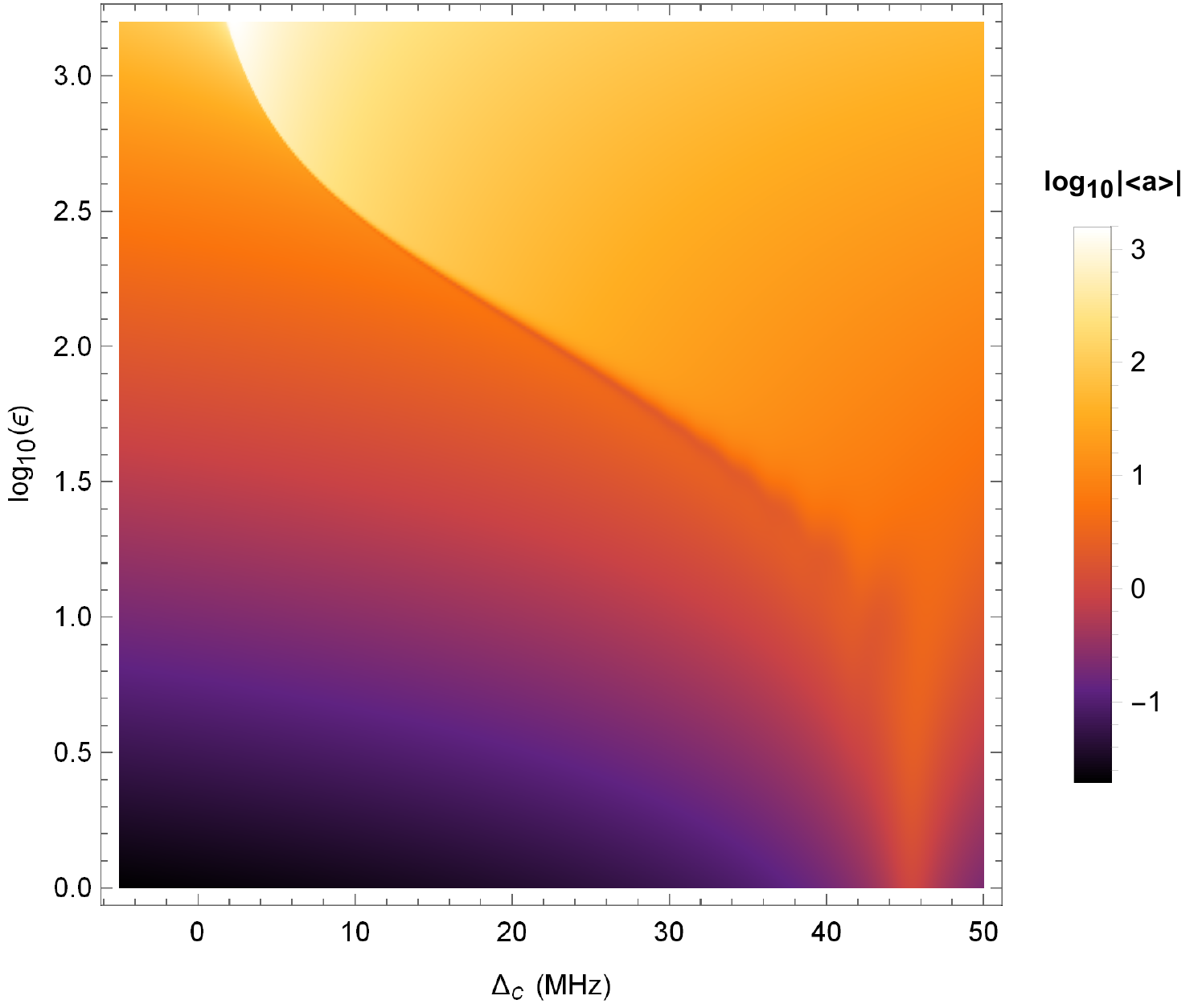}
    \caption{(Colour Online) Plots of cavity reflection $R$ and $|\expect{a}|$ as a function of the detuning $\Delta_c$ of the drive from the bare cavity frequency and drive amplitude $\epsilon$ when cavity and qubit are coupled in the strong dispersive regime. Other parameters are $\Delta_{ct}/2\pi=\SI{2.5}{\giga\hertz}$, $g/2\pi=\SI{340}{\mega\hertz}$, $\chi/2\pi=\SI{-220}{\mega\hertz}$, $\gamma_c/2\pi=\SI{2}{\mega\hertz}$ and $\gamma_t/2\pi=\SI{0.1}{\mega\hertz}$. At lower power, multiple dips in the reflection (less visible peak in the transmission) are visible, corresponding to the cavity frequency changing as a function of transmon state occupation. The reflection dip shifts towards the cavity resonance as the power is increased, approaching it asymptotically at very high power. The same shift of the cavity resonance is seem in the transmission spectrum,  in agreement with recent experimental results \cite{Reed2010}.}
    \label{fig:dispersive2}
\end{figure}

\subsection{Recovering the Cavity Moments}

In a typical experimental setup with a qubit interacting with the electromagnetic field of a 2D or 3D superconducting cavity, the most accessible measurements that can be performed are reflection from or transmission through the cavity. We therefore wish to calculate the moments of the cavity mode from those we have calculated for the transmon. To do this we return to the relations in Eq. (\ref{eqn:elim}), which were used to eliminate the cavity, and use these to write the cavity moments in terms of the transmon moments. This process is outlined in Appendix \ref{app:cavmom}. The first two such relations are

\begin{equation}
\expect{a} = \frac{2}{\tilde{\gamma}_c}(\epsilon -g\expect{b}),
\end{equation}
\begin{equation}
\expect{a^\dag a} = \frac{4}{|\tilde{\gamma}_c|^2}(|\epsilon|^2 - g\epsilon^* \expect{b} -g \epsilon \expect{b}^* +g^2 \expect{b^\dag b}).
\end{equation}
The amplitude of the field emitted from the cavity is proportional to $\expect{a}$. In addition we can plot the amplitude of the reflected field $R$, normalised by the drive strength. This is commonly measured in experiments where the cavity has only a single port and is given by
\begin{equation}
    R= \left|1-\frac{\gamma_c\expect{a}}{\epsilon}\right|.
\end{equation}

As the result in Eq. (\ref{eqn:moments}) and therefore expressions for $\expect{a^{\dag m} a^n}$ are analytic, it is possible to plot values of all moments over very large ranges of parameter space and in particular over many orders of magnitude of drive power, allowing us to explore regimes where the cavity is highly populated and simulation is unfeasible.

\subsection{Transmon Spectra}
\label{sec:spectra}
A standard driven quantum Duffing oscillator with nonlinearity $\chi$ will display evenly-spaced transmission peaks when driven at $\omega_r+k\chi$, for all positive integers $k$, where $\omega_r$ is the resonator frequency. In a frame rotating at the drive frequency this will correspond to $\Delta_c=k\chi$. We generalise this notion to predict the location of peaks in the transmon excitation for our combined system. Taking Eq. (\ref{eqn:FPE1}), we can work backwards to obtain an effective Hamiltonian for the transmon, after the cavity has been eliminated,
\begin{multline}
    H_t= \left[\Delta_c + \Delta_{ct} - \frac{4g^2 \Delta_c }{\gamma_c^2 + 4 \Delta_c^2} \right] b^\dag b + \frac\chi2 b^\dag b^\dag b b
    \\
    +\frac{2g}{\gamma_c + 2i \Delta_c} b^\dag + \frac{2g}{\gamma_c - 2i\Delta_c} b, 
\end{multline}
where we have written $\Delta_t = \Delta_c+ \Delta_{ct}$, with $\Delta_{ct}$ the cavity-transmon detuning.  In addition the effective decay rate for the transmon is $\gamma_t + 4g^2\gamma_c/(\gamma_c^2 + 4\Delta_c^2)$, which is consistent with the Purcell effect of coupling to the cavity. We predict peaks will occur at
\begin{equation}
    \Delta_c + \Delta_{ct} - \frac{4g^2 \Delta_c}{\gamma_c^2 + 4 \Delta_c^2}=k\chi , k \in \mathbb{Z}^+,
\end{equation}
which in fact holds exactly in all cases we plot. The higher order peaks require the transmon and cavity to be more significantly excited and therefore will appear at higher powers, but this model does not tell us at what drive strength they will appear. The actual device response is therefore strongly dependant on the drive power. For each value of $k$ there are three difference solutions for $\Delta_c$, suggesting that, in general the system behaves like three different non-linear oscillators in three distinct regions of of the drive frequency space.

In the case that the cavity and transmon are resonant the $k=0$ solutions can be expressed simply as $\Delta_c=0,\pm\sqrt{g^2-\gamma_c^2/4}$. In the strong coupling limit $g \gg \gamma_c$, this gives rise to the well known vacuum Rabi splitting of the cavity resonance \cite{Blais2004}. In Fig. \ref{fig:resonant} we show the cavity spectrum as a function of frequency and power. In the resonant regime we see that there is almost no transmission at the bare cavity frequency, with two peaks separated by $2g$ at low power. As the drive strength increases, each peak splits into two, displaying the supersplitting described in \cite{Bishop2008}. Transitions between higher cavity-transmon states then also appear at higher powers, with the nonlinearity increasing as higher levels are occupied. At very high powers there is a single bright peak at the bare cavity frequency as the drive overcomes the nonlinearity of the transmon. This behaviour is predicted by the Jaynes-Cummings model and \cite{Bishop2010} and seen in experiments \cite{Fink2010}. Despite the fact that the eigenstates of the system in this regime are strongly mixed between the cavity and transmon, and the vacuum Rabi splitting is caused by the exchange of excitations between atom and cavity, these features of the steady state behaviour all survive the adiabatic elimination procedure.

In the strong-dispersive regime $g^2/\Delta_{ct} > \gamma_{c,t}$, which is generally considered more relevant for quantum information processing, the system behaves differently depending on if it is driven near the bare cavity of bare qubit frequencies. Near the transmon frequency, as shown in Fig. \ref{fig:dispersive1}, the system behaves like a quantum Duffing oscillator with a dispersively shifted fundamental frequency of approximately $-\Delta_{ct}-g^2/\Delta_c$, and peaks separated by $\chi$. These peaks correspond to the transitions between adjacent levels of the transmon. Near the bare cavity frequency, the oscillator behaves as though it possesses a different nonlinearity, which decreases the more the transmon in populated (see Fig. \ref{fig:dispersive2}). Again, the fundamental frequency is dispersively shifted at approximately $g^2/\Delta_{ct}$. At low power, there are several resolvable transmission peaks, which correspond to the dependence of the cavity frequency on the occupation of the first few transmon energy levels. As the power is increased, these peaks can not longer be resolved and a single transmission peak forms which shifts towards the bare cavity frequency.  At high powers, the system behaves like a linear oscillator very close to the bare cavity resonance, as is observed experimentally  \cite{Reed2010,Paik2011,Bishop2010}. 

The reflection spectrum of the system mirrors many of the features of the transmission, displaying multiple distinct peaks at moderately low powers, corresponding to the position of the cavity resonance shifting as a function of the number of excitations in the transmon. In a recent paper it has been shown that at low powers our solution agrees well with both experimental reflection data and full master equation simulations \cite{Themis2016}. As the power increases, this become a single reflection dip which sweeps towards the bare cavity frequency. If a non-zero temperature environment is considered, then there will be some excited state population even for zero drive and we expect that these dips would appear at lower powers.

In reality the transmon possesses a cosine potential \cite{Koch2007}, which is not well approximated by our Duffing oscillator model for all energy levels and, we must therefore consider this when interpreting our results. The quartic approximation is appropriate only for those levels which are contained within the cosine potential wells, which vary in number depending on the ratio of the Josephson and charge energy $E_J/E_C$ for the specific device. For typical devices this is the first four to eight excited states of the device \cite{Braumuller2015,Peterer2015}. Almost all of the features we describe above, for both the resonant and dispersive regimes occur in the regime where we expect the Duffing model to hold. Only at very high powers, when the transmission peak is returning to the bare cavity frequency and becomes very bright, do we expect higher transmon levels to become relevant. We discuss the applicability of the Duffing model further in Appendix \ref{app:model} in addition to plotting $\expect{b^\dag b}$ to illustrate where we expect the model to break down.

\begin{figure}
    \centering
    \includegraphics[width=\columnwidth]{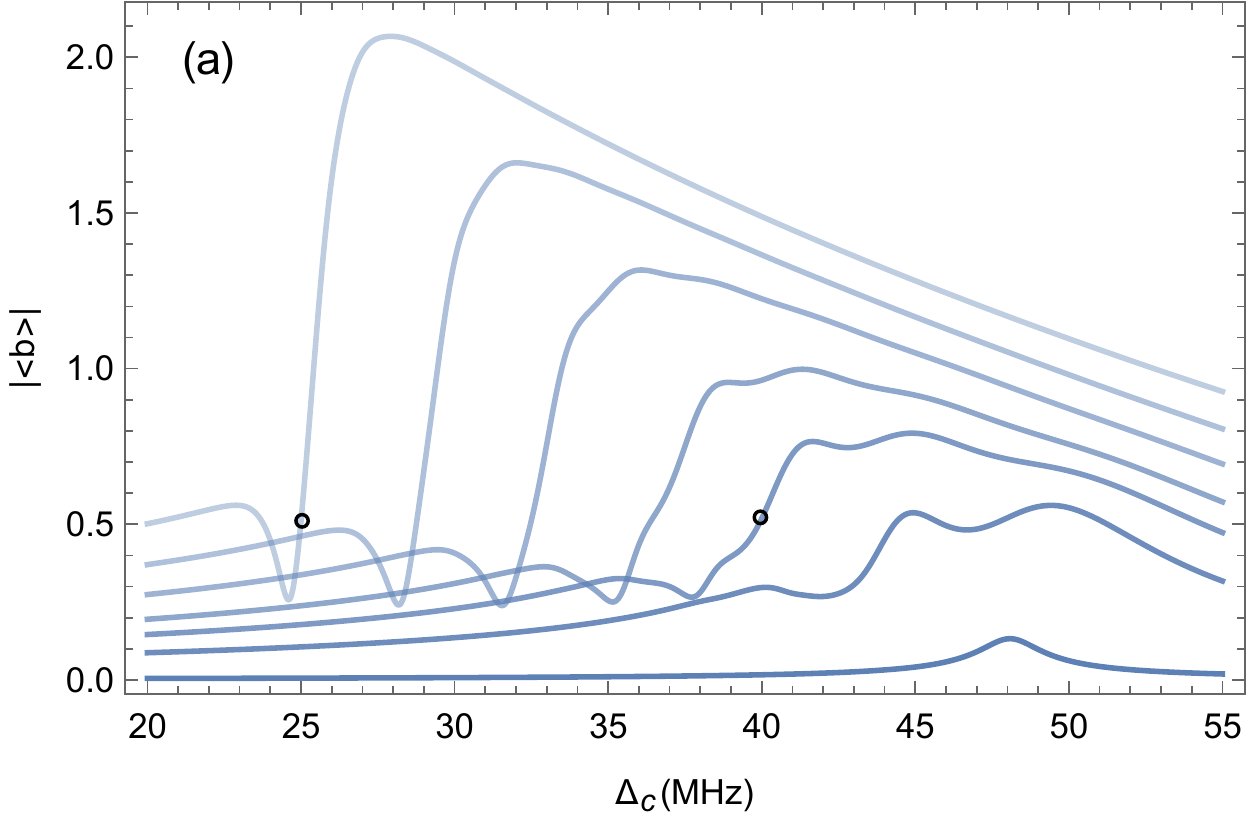}
    \includegraphics[width=.49\columnwidth]{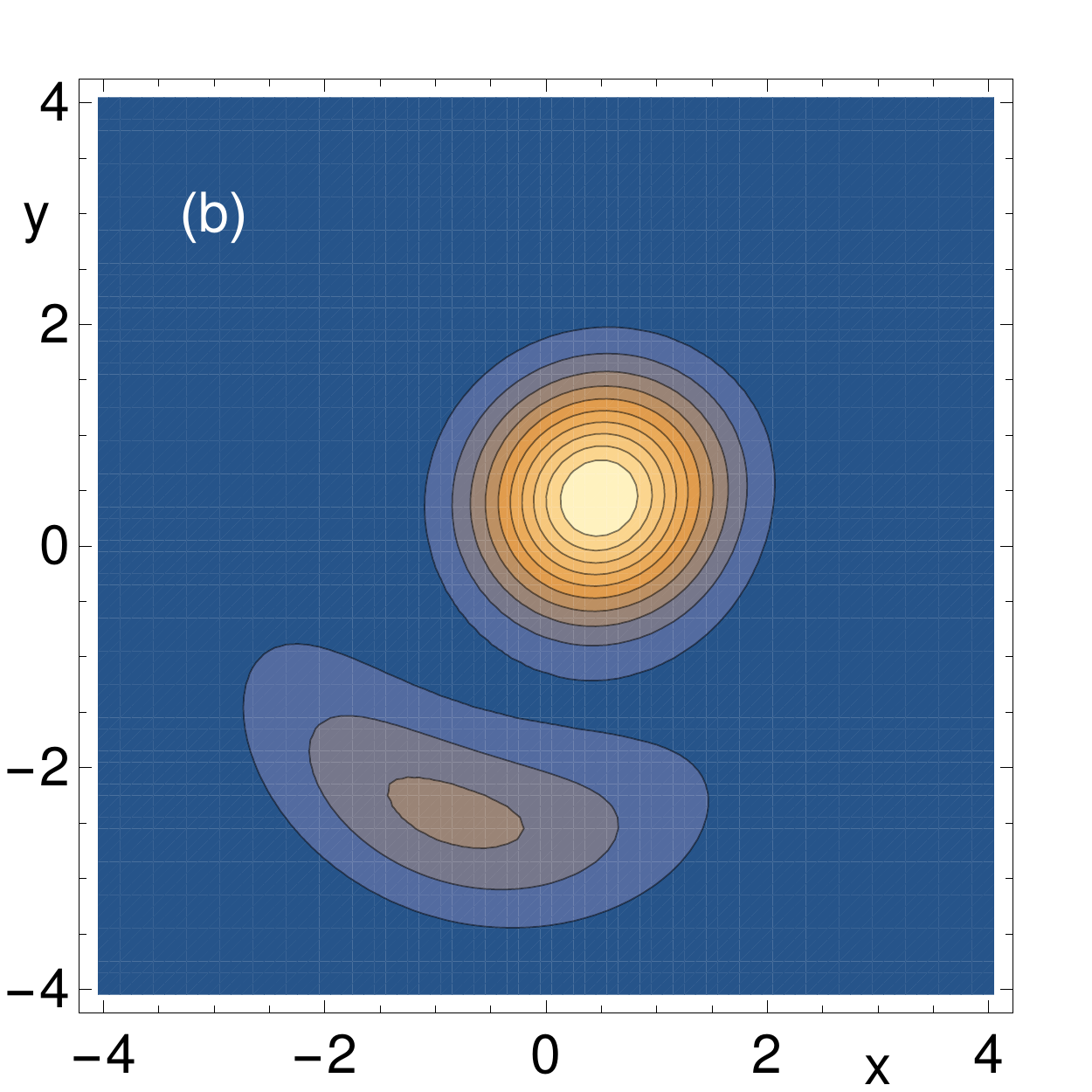}
    \includegraphics[width=.49\columnwidth]{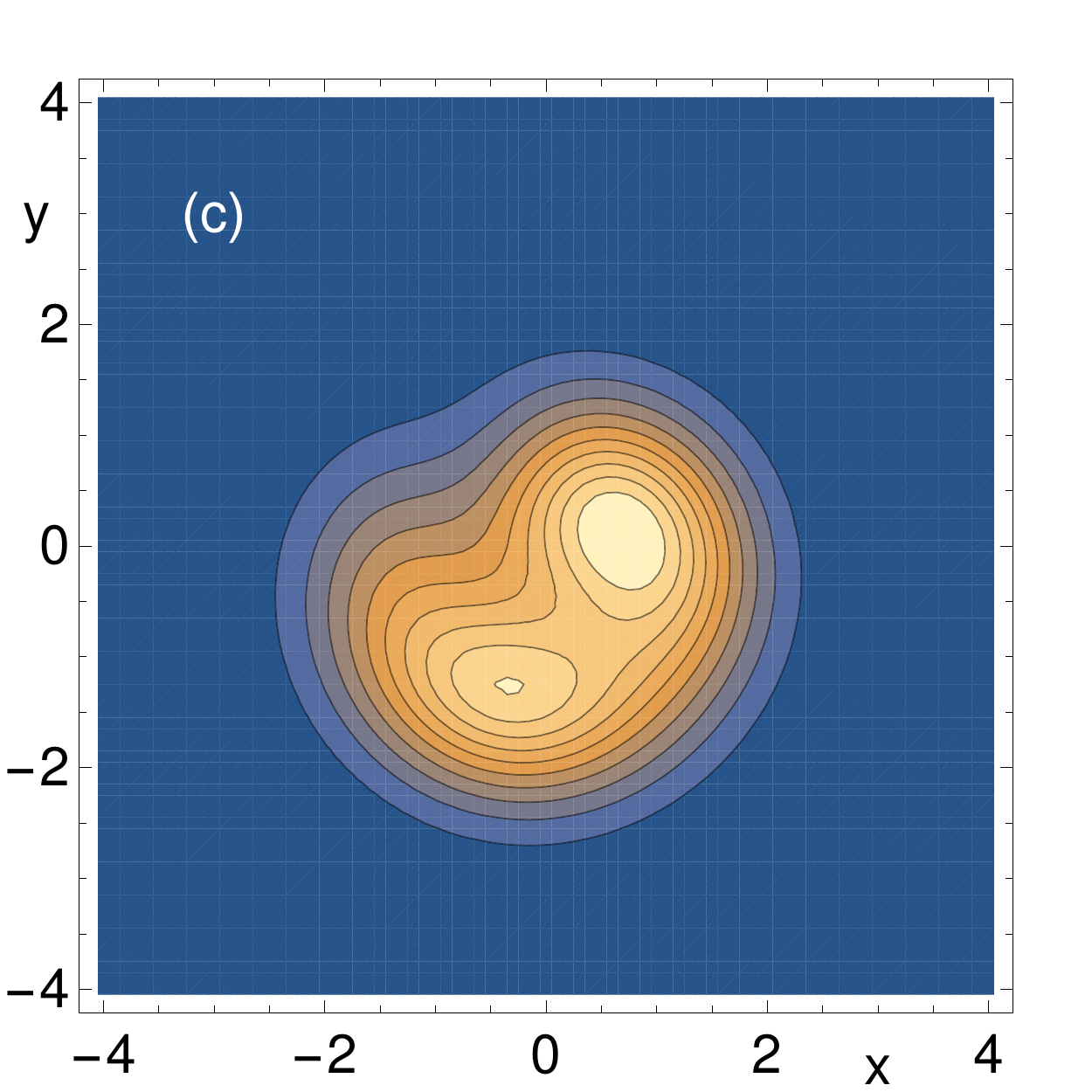}
    \caption{(Colour Online) (a) Plot of transmon field amplitude $|\expect{b}|$ as a function of drive detuning from the bare cavity $\Delta_c$, plotted for various values of the drive amplitude $\epsilon$ for a cavity transmon system with parameters $\Delta_{ct}/2\pi=\SI{2.5}{\giga\hertz}$, $g/2\pi=\SI{350}{\mega\hertz}$, $\chi/2\pi=\SI{-220}{\mega\hertz}$, $\gamma_c/2\pi=\SI{2}{\mega\hertz}$ and $\gamma_t/2\pi=\SI{0.1}{\mega\hertz}$. Values of $\epsilon/2\pi$ (from darkest to lightest) are \SI{1}{\mega\hertz}, \SI{18}{\mega\hertz}, \SI{30}{\mega\hertz}, \SI{40}{\mega\hertz}, \SI{56}{\mega\hertz}, \SI{75}{\mega\hertz} and \SI{100}{\mega\hertz}. (b) $Q$-function of the transmon field with $\epsilon/2\pi=\SI{30}{\mega\hertz}$ and $\Delta_c/2\pi=\SI{40}{\mega\hertz}$ and (c) $Q$-function of the transmon field with $\epsilon/2\pi=\SI{100}{\mega\hertz}$ and $\Delta_c/2\pi=\SI{25}{\mega\hertz}$. These two points are marked with black circles in (a). We see that, in addition to the bifurcation of the cavity, the model predicts bistability for the qubit when the system is driven near to the cavity resonance.  At higher powers the behaviour of the system becomes very similar to that of the  standard quantum Duffing oscillator with the characteristic dip due to coherent cancellation of the two steady states. The frequency at which the transmon (and cavity) bifurcation occurs shifts towards the bare cavity frequency as the power is increased, as seen in Fig. \ref{fig:dispersive2} for the cavity.  A low power, the transmon field response splits into several peaks, corresponding to transmission peaks of the cavity at different transmon occupation numbers, but bistability can still be seen in the transmon $Q$-function.}
    \label{fig:dip1}
\end{figure}

\subsection{Transmon Bistability}
Plots of the transmon $Q$-function allow us to study additional features of the oscillator state. In particular, a bimodal $Q$-function is indicative of bistability in the steady state, with switching occurring due to tunnelling between the two states \cite{Lin2015}. In our model we see that, when the system is driven near the cavity resonance at sufficient power, a bistability occurs simultaneously for both the cavity and transmon fields.  This is different to the Duffing-type behaviour of the cavity in the lower power regime, where it is possible to consider the qubit as providing only a small nonlinear perturbation to the cavity field. In Fig. \ref{fig:dip1} we show that the characteristic dip in $|\expect{b}|$, corresponding to the coherent cancellation of the two steady states with opposite phases, can be seen in the transmon field at high powers. The form of $|\expect{b}|$ as a function of $\Delta_c$ looks identical to the quantum Duffing oscillator \cite{Drummond1980}, with the dip shifting towards the bare cavity frequency as the power is increased. At very high powers, when the dip has shifted to the cavity frequency, this dip stops being present as the whole system begins to behave linearly. At lower powers, we see multiple peaks in the transmon occupation, corresponding to the peaks in the cavity field seen in Fig. \ref{fig:dispersive2}, which arise from the dependence of the cavity frequency on the transmon occupation. Even though the dip can non longer be seen at such powers, the bistability still persists and can be clearly seen in the transmon $Q$-function.

\section{The parametrically driven Duffing oscillator}

Our second system is a single Duffing oscillator which is driven both parametrically and coherently. Parametrically driven oscillators have been studied extensively in circuit QED for applications including squeezing generation \cite{Yurke1988} and qubit readout \cite{Lin2014,Wustmann2013}. The parametrically driven Duffing model has also been investigated more fundamentally, including switching rates near bifurcation points \cite{Dykman2012, Lin2015}, critical exponents of the phases transition \cite{Dykman2007} and metastable lifetimes of the steady state \cite{Drummond1981}. The Hamiltonian of the system is

\begin{multline}
\label{eqn:H2}
H_2= \omega_r c^\dag c +i(\epsilon_1  e^{-i\omega_{d_1} t} c^\dag - \epsilon_1^* e^{i\omega_{d_1} t}c )
\\
 + \frac i2(\epsilon_2 e^{-i\omega_{d_2} t}c^\dag c^\dag  - \epsilon_2^* e^{i\omega_{d_2} t} c c) +\frac U2 c^\dag c^\dag c c,
\end{multline}
where $c$ is the annihilation operator for the resonator mode, $\omega_r$ is the resonator frequency, $\epsilon_1$ and $\epsilon_2$ encode the amplitude and phases of the coherent and parametric drives respectively and $U$ is the strength of the quartic nonlinearity of the system. In order that this system can be cast in time independent form, we require that $\omega_{d_2}=2\omega_{d_1}$. In this case we can transform into a rotating frame at the drive frequency with the Hamiltonian

\begin{multline}
\tilde{H}_2= \Delta c^\dag c +i(\epsilon_1 c^\dag - \epsilon_1^* c) + \frac i2(\epsilon_2 c^\dag c^\dag - \epsilon_2^* c c)
\\
+\frac U2 c^\dag c^\dag c c, 
\end{multline}
where $\Delta=\omega_r-\omega_{d_1}$ is the detuning of the two drives from the cavity frequency. Additionally, we account for single photon loss at rate $2\gamma_1$ and the loss of pairs of photons at rate $\gamma_2$, so that the master equation for the system is given by
\begin{equation}
\dot{\rho} = -i[\tilde{H}_2,\rho] +  \mathcal{L}[\sqrt{2\gamma_1}c]\rho  + \mathcal{L}[\sqrt{\gamma_2}cc]\rho \end{equation}
The FPE for this system can then be easily written down using the standard rules, producing
\begin{multline}
\frac{\partial P_2(\alpha,\beta)}{\partial t}= \biggl[-\pd{\alpha}[\epsilon_1 - \kappa_1 \alpha +  (\epsilon_2 -\kappa_2 \alpha^2) \beta]
\\
- \pd{\beta}[\epsilon_1^*- \kappa_1^*\beta + (\epsilon_2^* -\kappa_2 ^* \beta^2)\alpha] 
\\
+ \frac 12 \spd{\alpha} (\epsilon_2- \kappa_2 \alpha^2) + \frac 12  \spd{\beta} (\epsilon_2^* -\kappa_2^* \beta^2) \biggr]P_2(\alpha,\beta),
\end{multline}
where $(\alpha,\beta)$ are the phase space coordinates of the resonator and we have defined $\kappa_1 = \gamma_1 + i\Delta$ and $\kappa_2=\gamma_2 + i U$. The solution to this system is of the form of that in \cite{Drummond1981}, but with the coefficient of the nonlinearity replaced by $\kappa_2$, which allows the strength of the nonlinearity which to be varied independently of the other parameters through $U$, and additionally includes the two photon loss. The moments of the oscillator can be written in terms of the hypergeometric function $\tensor[_2]{F}{_1}$ and are given by
\begin{equation}
\expect{c^{\dag m }c^n}=\frac{I_{mn}}{I_{00}},
\end{equation}
with
\begin{multline}
\label{eqn:ampmoms}
I_{nm}=  \sum_{j=0}^{\infty}\frac{2^j}{j!}\left(-\sqrt{\frac{\epsilon_2}{\kappa_2}}\right)^{j+m} \left(-\sqrt{\frac{\epsilon_2^*} {\kappa_2^*}}\right)^{j+n}
\\\tensor[_2]{F}{_1}(-j-m,A-B,2A;2)\tensor[_2]{F}{_1}(-j-n,A^*-B^*,2A^*;2),
\end{multline}
where we have defined two constants $A=\kappa_1/\kappa_2$,  and $B=-\epsilon_1/\sqrt{\epsilon_2 \kappa_2}$. As with the cavity-transmon system, it is also possible to derive exact expressions for $P(n)$ and the $Q$-function, which are of a similar form and are given in Appendix \ref{app:pandq}. 

\begin{figure}
  \centering
    \includegraphics[width=.9\columnwidth]{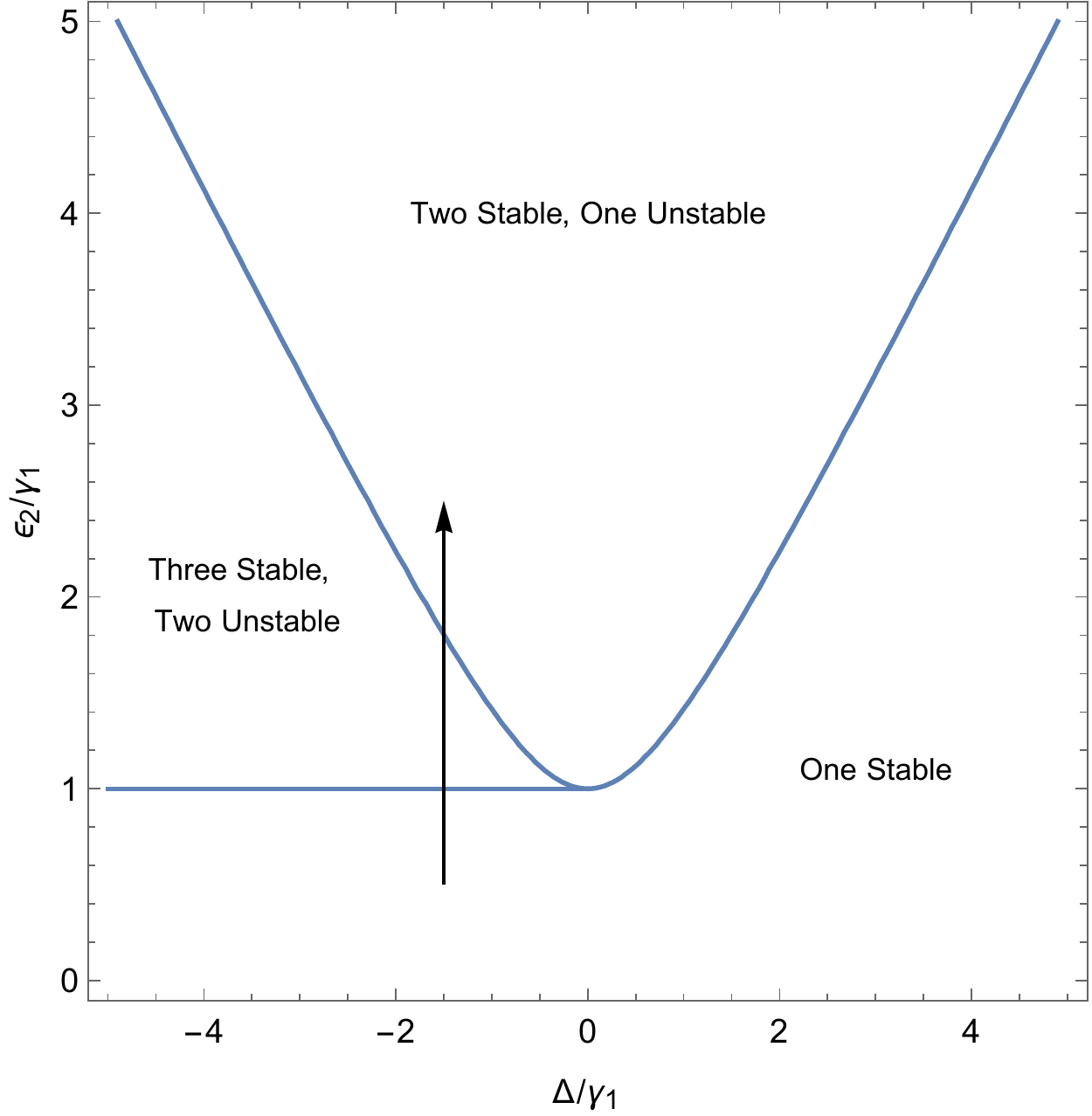}

\caption{(Colour Online) Classical phase diagram of the parametrically driven Duffing oscillator in the $(\Delta,\epsilon_2)$ plane based on number of fixed points. The plane is dived into 3 regions by the lines $\epsilon_2^2 - \Delta^2=\gamma_1^2$ and $\epsilon_2=\gamma_1$, corresponding to the well-known threshold of the parametric oscillator. The system has up to five fixed points, of which up to three can be stable. The classical diagram is unaffected by the value of $U$, which only modifies the amplitude, and therefore separation of the solutions. Figs. \ref{fig:trans1} and \ref{fig:trans2} show the evolution of the steady state as the phase space is traversed parallel to the arrow, but for larger negative detunings}
\label{fig:phases}
\end{figure}

\subsection{Mean-field phases}
In the case where $\epsilon_1=\gamma_2=0$, it is simple to solve a classical mean field equation of motion for the steady state of this system
\begin{equation}
\frac{\partial \alpha}{\partial t}=\epsilon_2\alpha^* - iU\alpha^2\alpha^* -\gamma_1 \alpha -i\Delta \alpha= 0.
\end{equation}

This system has up to 3 solutions for the amplitude: $\alpha=0$ and
\begin{equation}
 |\alpha|^2  = \frac{- \Delta \pm \sqrt{|\epsilon_2|^2-\gamma_1^2}}{U}.
\end{equation}
Solving for the phase shows that these solutions come in pairs with opposite phases. Additionally, the stability of these fixed points can be determined by finding the eigenvalues of the Jacobian matrix of the system \cite{Hirsch2004}. This allows us to divide the $(\Delta, \epsilon_2)$ plane into 3 distinct phases based on the numbers of solutions at each point in parameter space, as shown in in Fig. \ref{fig:phases} \cite{Zorin2011,Lin2015}. Classical phases with one, two and three stable states exist, with the boundary between the one- and two-solution phases appearing in the same place as the threshold of an ideal parametric amplifier. The existence of the non-linearity $U$ does not affect the structure of the classical phase diagram, but does reduce the amplitude of the steady states.

\begin{figure}
  \centering
    \includegraphics[width=\columnwidth]{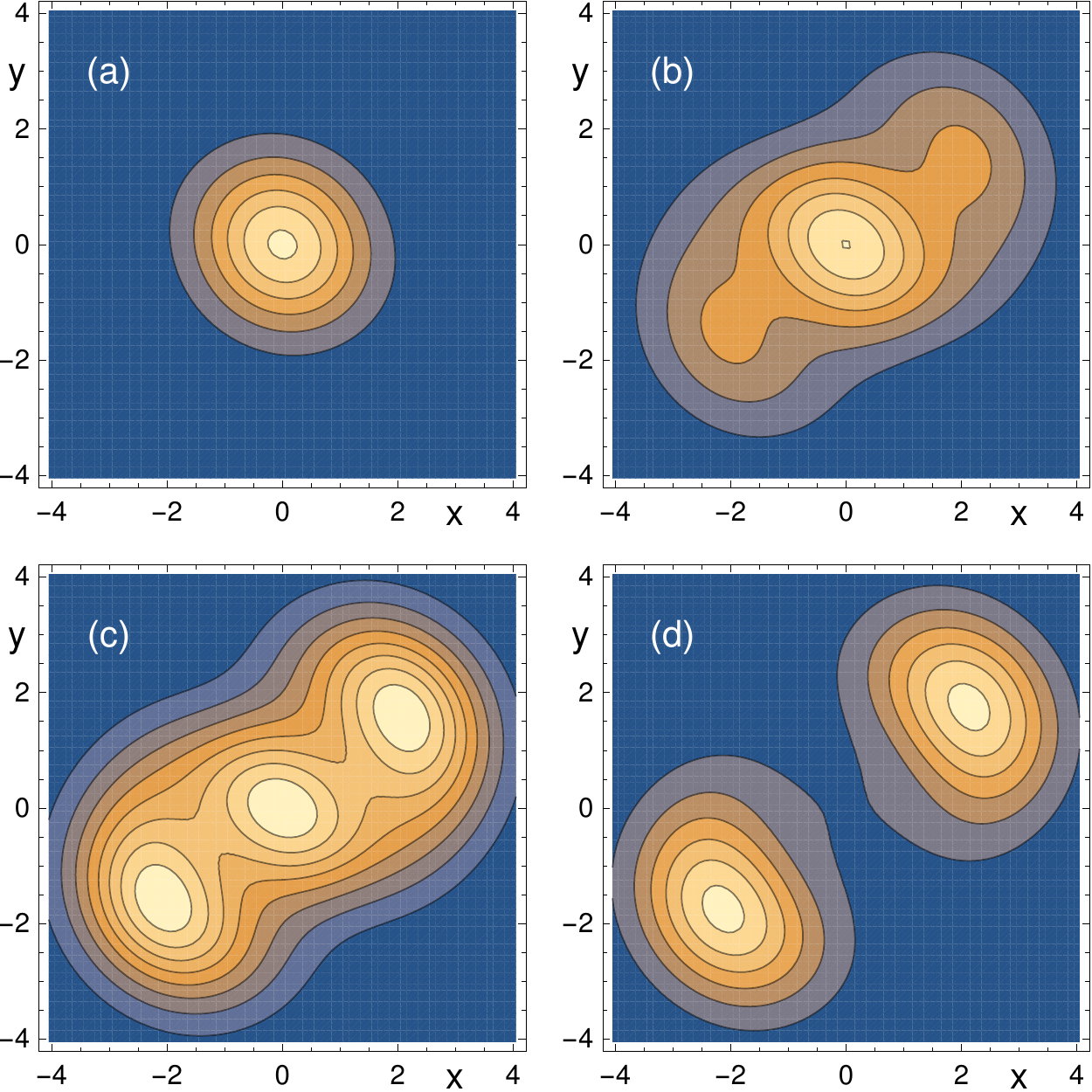}

\caption{(Colour Online) Analytical $Q$-function plots for a parametrically driven Duffing oscillator with $\gamma_1/2\pi=\SI{1}{\mega\hertz}$, $U=5\gamma_1$ and $\Delta=-12 \gamma_1$ driven at four different drive strengths (a) $\epsilon_2=2\gamma_1$ (b) $\epsilon_2=4.25\gamma_1$ (c) $\epsilon_2=4.75\gamma_1$  and (d) $\epsilon_2/2\pi=\SI{6.25}{\mega\hertz}$. Over this range of drives, the resonator state crosses two classical phase boundaries and we see the emergence of three stable points, followed by only two. The addition of a significant $U$, makes the threshold at $\epsilon_2=\gamma_1$ appear much later than when $U=0$ as it is harder to add photons to the resonator, while the second boundary seems to appear much earlier than predicted, as there is extremely low probability of being in the $\alpha=0$ state in much of the phase.}
\label{fig:trans1}
\end{figure}

\begin{figure}
  \centering
    \includegraphics[width=\columnwidth]{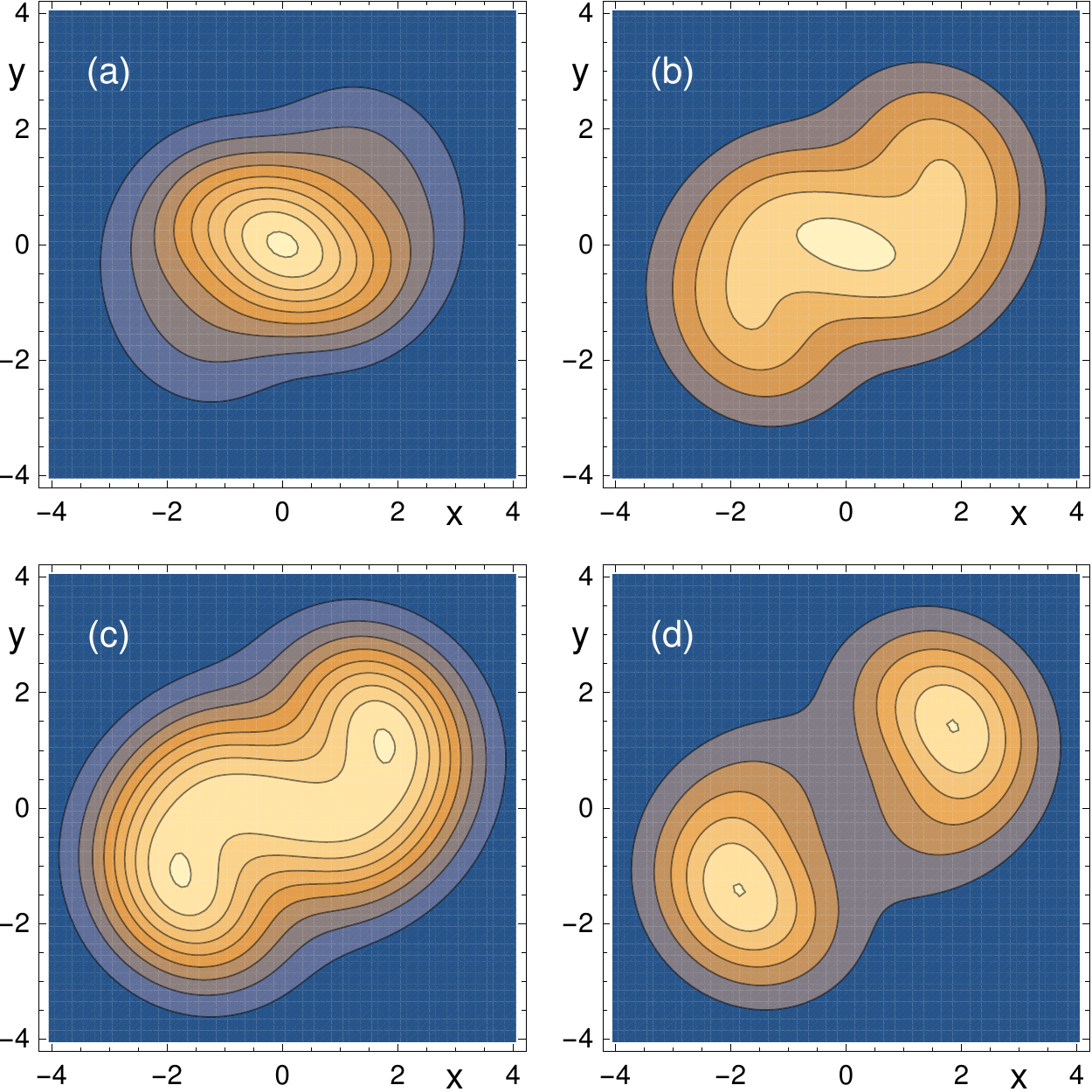}

\caption{(Colour Online) Analytical $Q$-function plots for a parametrically driven Duffing oscillator with $\gamma_1/2\pi=\SI{1}{\mega\hertz}$, $U=5\gamma_1$ and $\Delta=-8\gamma_1$ driven at four different drive strengths (a) $\epsilon_2=2.75\gamma_1$ (b) $\epsilon_2=3.5\gamma_1$ (c) $\epsilon_2=3.75\gamma_1$ and (d) $\epsilon_2=5\gamma_1$. At a smaller detuning than in Fig. \ref{fig:trans1}, the effect of a large $U$ is to prevent all of the stable states from being resolved. While the system crosses two classical phase boundaries, we only see a single fixed point, which eventually become bistable at high enough driving powers.}
\label{fig:trans2}
\end{figure}

\subsection{Phase transitions in the quantum system}
In the full quantum system, the hard phase boundaries of the classical system are not present, and analytical $Q$-functions allow us to to study how these states develop as the classical boundary is crossed. The regime that is of particular interest is where $U\gg \gamma_1$, firstly because as $U \to 0$, the system reverts to the ideal degenerate parametric amplifier, but also because the presence of the nonlinearity resists the addition of excitations to the system. This means that the stable states of the system are kept closer together in phase space, allowing multistabilities of the quantum system to be more easily observed and preventing the system from behaving classically. In Figs. \ref{fig:trans1} and \ref{fig:trans2} we plot $Q$-functions for increasing drive strength for a fixed value of $U=5\gamma_1$ and two values of the detuning $\Delta=-8\gamma_1,-12\gamma_1$. For the larger detuning, we see all three phases manifest themselves. The first phase transition, from a single stable point to three, occurs later than predicted classically due to the nonlinearity, while the transition from three to two stable points seems to occur earlier, as while there is a probability of being in the $\alpha=0$ state, it is extremely small for much of the phase.

When the drive is less detuned from the cavity frequency, the separation of the fixed points is smaller and therefore we only see two distinct phases in the resonator $Q$-functions and the state appears to move directly from one fixed point to two, without every clearly seeing three. When we include a small classical drive ($\epsilon_1>0$), we see that, for both values of the detuning, that the steady state is pushed towards either of the non-zero amplitude fixed points, depending on the phase of the signal, with the probability of being found in the other states reducing. Controlling this type of transition has recently been studied by another group \cite{Bartolo2016}. For a sufficiently large signal the resonator will always be found in a coherent steady state. It is therefore possible to use this system in the three-stable point phase as a detector of small coherent signals, which forms the basis for proposed period-doubling bifurcation detectors \cite{Zorin2011}.

\section{Generation of squeezing}

\begin{figure}
    \centering
    \includegraphics[width=\columnwidth]{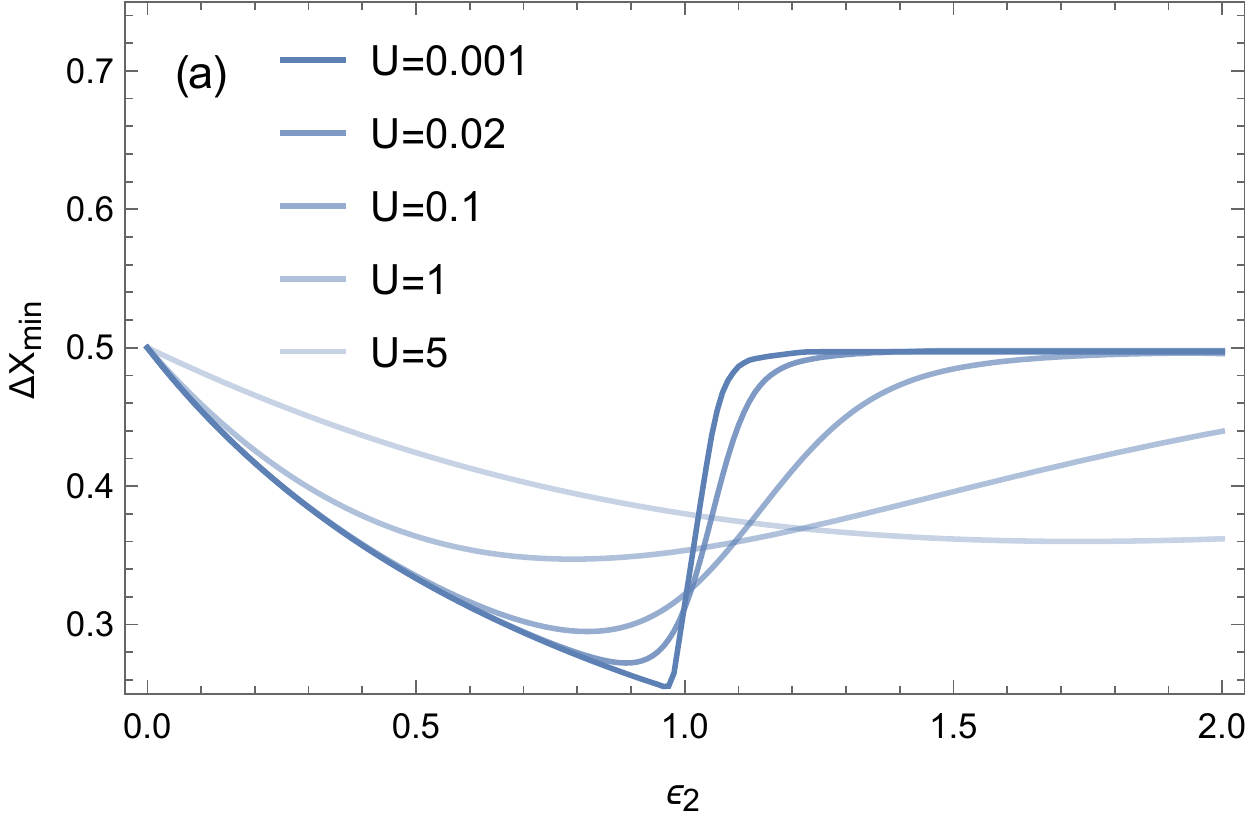}
     \includegraphics[width=\columnwidth]{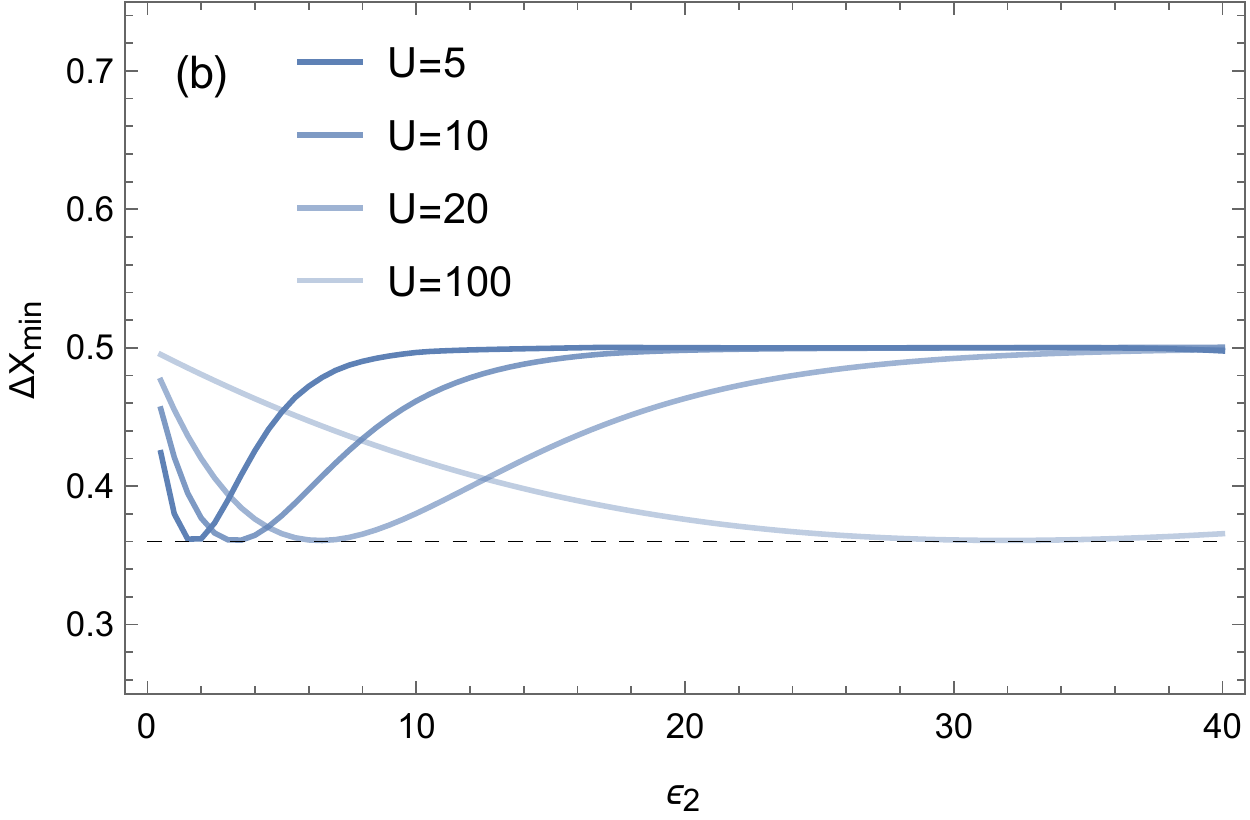}
    \caption{(Colour Online) Minimum quadrature uncertainty for a non-ideal degenerate parametric amplifier as a function of parametric drive $\epsilon_2$ for different values of the nonlinearity $U$ (given in MHz). Dissipation occurs at a rate $\gamma_1=\SI{1}{\mega\hertz}$. (a) For very small nonlinearities, the system behaves like an ideal degenerate parametric amplifier with a sharp threshold at $\epsilon_2=\gamma_1$ where the minimum quadrature uncertainty goes to $1/2$ of that of the vacuum, corresponding to an uncertainty $\Delta X_{min}=0.25$. As $U$ is increased, this threshold is moved initially slightly lower, and then to higher powers, while the maximum squeezing that can be achieved is reduced. Past the minimum there is a period where the steady state has bifurcated, but the uncertainty in some direction is still less than than of the vacuum. (b) For $U \gg \gamma_1$, $\Delta X_{min}$ has a fixed minimum value at around $0.36$ (dashed line) and the position of the minimum is at approximately $U/3$. For very large drives the state is made up of two well-separated coherent states, each with the same uncertainty as the vacuum.}
    \label{fig:squeezing}
\end{figure}

When driven below threshold and on resonance, in the phase with a single steady state, the system can behave as a degenerate parametric amplifier and produces squeezing of the resonator state. Generation and measurement of squeezing has been the subject of much recent research in the field of circuit QED \cite{Zagoskin2008,Didier2014,Boissonneault2014,Elliott2015}. When $U=0$, it is known that the maximum squeezing the can be achieved is a factor of 2, reducing the fluctuations in one field quadrature to 50\% of those of the vacuum state \cite{Milburn1981}. A complete treatment of the parametric down conversation process that includes both modes and then eliminates the pump mode, introduces a small quartic term, but a nonlinearity could also be introduced, for example, by the presence of a Josephson junction, or a dispersively coupled qubit. The strong coupling that is possible in circuit QED when compared with most systems in the optical regime means that this nonlinear term can in principle be very large. The nonlinearity has the potential to limit the degree of squeezing that can be achieved, while also shifting the threshold due to $U$ resisting the addition of excitation to the system, as discussed above. A reduction in the squeezing of the internal field, will also lead to a corresponding fall in the squeezing of the emitted field. 

The degree of squeezing present in the cavity field can be characterised by the uncertainty in the field quadratures. Specifically we use the minimum uncertainty
\begin{equation}
\Delta X_{min} = \min_{\theta \in [0,\frac\pi2]} \left(2\expect{c^\dag c} + e^{2i\theta}\expect{cc} + e^{-2i\theta}\expect{c^\dag c^\dag} +\frac12  \right),
\end{equation}
where $\theta$ determines the direction in phase space that the uncertainty is measured in. In Fig. \ref{fig:squeezing}, we show the minimum quadrature uncertainty as a function of drive strength for nonlinearities that range from much smaller than the dissipation to many times greater. While our solutions for the moments is not defined for $U=0$, we can produce a plot for $U=0.001\gamma_1$, where the nonlinearity is insignificant compared with the dissipation, and see that the maximum squeezing comes very close to the ideal value of 0.25. We see that even a very small nonlinearity of $U=0.02\kappa$ causes a significant increase in the minimum uncertainty, and that this damage to the squeezing increases as $U$ approaches $\gamma_1$. Once $U \gg \gamma_1$, however, this trend stops. Even for very large nonlinearities, it is always possible to achieve a small amount of squeezing. The minimum quadrature uncertainty tends towards 0.36, and does not reduce further as the nonlinearity strength increases.

As in the previous section, increasing $U$ modifies the where the classical threshold of the parametric amplifier appears. This effect can be clearly seen in Fig. \ref{fig:squeezing}. For each value of $U$ there is a minimum in $\Delta X_{min}$ as a function of $\epsilon_2$. Below this minimum, the state is an ideal Gaussian squeezed state, while above it the state is bimodal, although it retains some degree of squeezing in one quadrature as this bifurcation occurs. The semiclassical treatment of this system places this threshold at $\epsilon_2=\gamma_1$, and we see that the behaviour of the quantum system as $U\to0$, tends towards a sharp jump in the uncertainty as the bifurcation occurs at this point. As $U$ is increased, the region over which this transition occurs is increasingly broadened, with the minimum uncertainty still occurs just as the bifurcation begins. Note that while plots of a particular field quadrature, such as those in \cite{Milburn1981}, show cusps in the uncertainty as this transition occurs, $\Delta X_{min}$ always varies smoothly. The initial effect of introducing a small $U$ is to lower the position of the threshold slightly, but it then rises as the nonlinearity resists the addition of photons to the resonator. For $U \gg \gamma_1$ the the threshold is at approximately $\epsilon_2=U/3$.

\section{Stabilisation of cat states}
\begin{figure}
    \centering
    \includegraphics[width=\columnwidth]{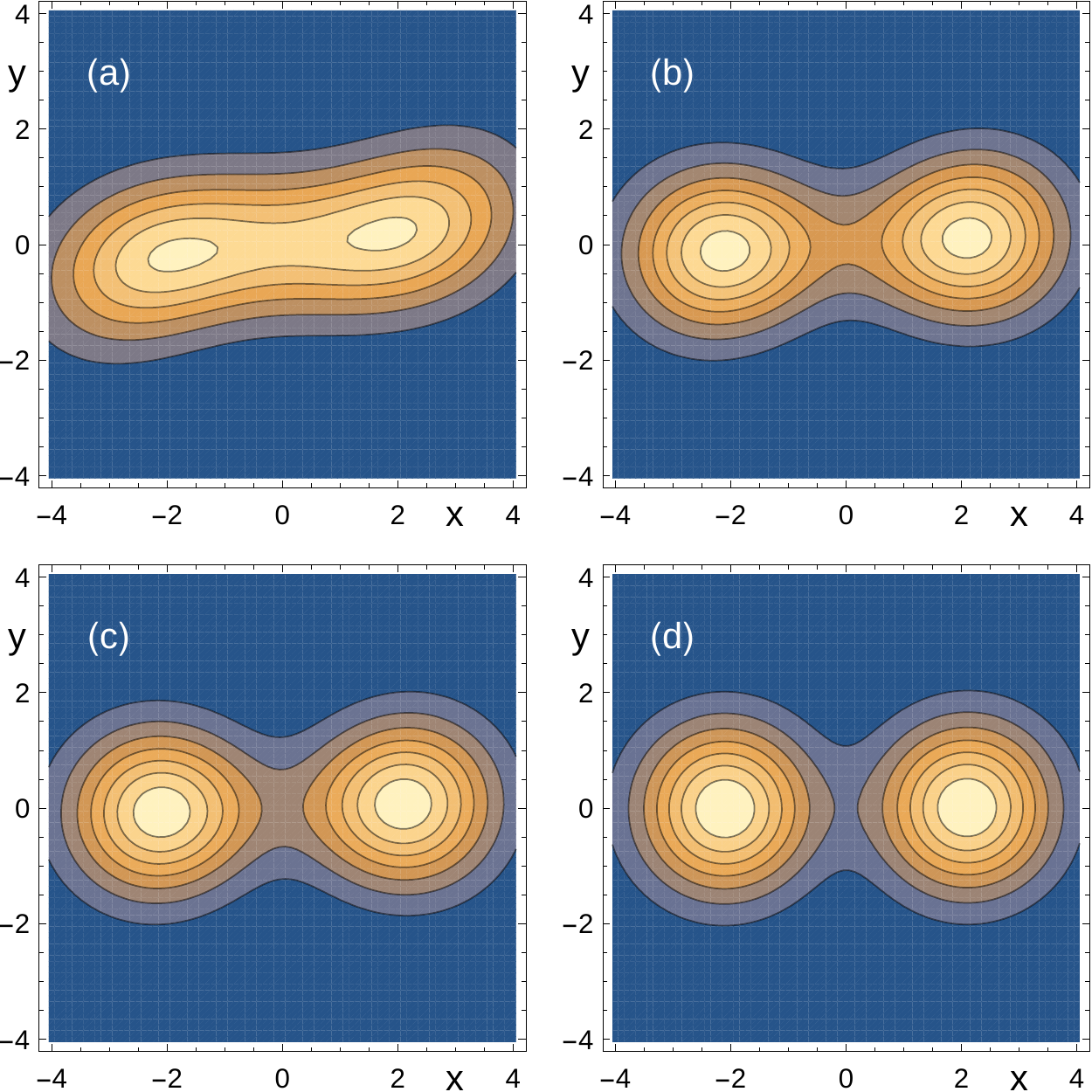}
    \caption{(Colour Online) $Q$-functions of nonlinear oscillator driven by a two photon process and with different ratios of single-photon to two-photon loss rates. The system has $U/2\pi=\SI{0.1}{\mega\hertz}$, $\gamma_1/2 \pi = \SI{1}{\mega\hertz}$ and the drive $\epsilon_2$ adjusted so that the average number of photons in the mode is 2.2. The other parameters are (a) $\gamma_1/\gamma_2=20$, $\epsilon_2/2\pi = \SI{1.15}{\mega\hertz}$ (b) $\gamma_1/\gamma_2=2$, $\epsilon_2/2\pi = \SI{2.28}{\mega\hertz}$ (c) $\gamma_1/\gamma_2=1$, $\epsilon_2/2\pi = \SI{3.4}{\mega\hertz}$ and (d) $\gamma_1/\gamma_2=.1$, $\epsilon_2/2\pi = \SI{23.5}{\mega\hertz}$. When single-photon loss is the dominant loss mechanism, the resonator nonlinearity causes distortions of the stable coherent states, reducing the fidelity of any stored state. Once $\gamma_2$ becomes comparable to $\gamma_1$, however, the distortions are reduced, with only a small `bridge' between the two states. Once the two-photon loss is much faster than the single photon loss the distortions are eliminated completely.}
    \label{fig:cats}
\end{figure}

Schr\"odinger cat states are a class of coherent state superpositions consisting of two coherent states with the same amplitude and opposite phase. These states can now be realised in circuit QED \cite{Vlastakis2013}. There is currently considerable interest in using theses state to store and process quantum information, taking advantage of the fact that cavity lifetimes are much longer than those of qubits  \cite{Mirrahimi2013,Ofek2016,Wang2016}. Storing information in these multiphoton states is also partially robust again the loss of single photons, whereas losing the excitation from a qubit will cause complete decoherence. Manipulation and read out of these cavity states is generally achieved via coupling to a superconducting qubit. In strong dispersive circuit QED it is common to perform an elimination of the qubit, producing an effective model of the form of Eqn. (\ref{eqn:H2}) with an $(a^\dag a)^2$ term \cite{Boissonneault2009}, known as the cavity self-Kerr. There is interest in using networks of such nonlinear cavities to perform quantum computation \cite{Goto2016, Goto2016a}. 

Recently, it has been demonstrated that driving a cavity parametrically via a four-wave mixing process, while simultaneously using this to remove pairs of photons from the resonator ($\gamma_2>0$), could be enable stabilisation of a cat state \cite{Leghtas2015}. This system has been studied using the positive $P$-representation \cite{Wolinsky1988}, showing that if $\gamma_1=0$ then all possible superpositions of the coherent steady states are themselves stable. If $\gamma_1>0$, then a recent paper has shown, by comparing analytical and master equation results, that the state eventually the superposition decays into a mixture of odd and even cat states, with single photon loss causing switching between the two \cite{Minganti2016}. A parity measurements can then be used to project the state back into the correct subspace. 

The presence of the Kerr nonlinearity in this system will distort the stabilised cat and reduce the fidelity of information storage. Even if $U$ is small, then this effect will become increasingly relevant as the combination of two-photon driving and parity measurements is used to preserve the state for many cavity lifetimes. This may lead to the need to increase the size of the cat to prevent overlap between the to states, increasing vulnerability to other loss mechanisms, for example via the qubit. A recent work showed that transient distortions in cat state preparation can be reduced using a two photon driving and a large $U$ in the presence of only single-photon loss \cite{Puri2016a}, but the phase information is still lost in the steady state. We investigate whether altering the ratio of one- and two-photon loss can alleviate distortions in the steady state. As the steady state of the system is mixed, the Wigner function is identical the the state Q-function, and there are no interference fringes, but the shape and overlap between the two coherent states can still tell us whether cats will be stabilised with good fidelity after the projective measurement

In Fig. \ref{fig:cats}, we plot $Q$-functions for the system for a constant $U$ and different values of $\gamma_2$, with $\gamma_1$ fixed and $\epsilon_2$ adjusted to keep the number of photons constant at 2.2. This size of cat is large enough that the overlap between the two coherent states is negligible in the ideal case \cite{Ofek2016}. We see that when the dominant source of energy loss is by single photons, there are significant distortions to the steady state and there is significant overlap between the two peaks, making it impossible to store information in the state. When the two rate are of comparable size, this overlap is already greatly reduced, with a small `bridge' in the $Q$-function between the two stable points, suggesting a small amount of switching between the two states. When $\gamma_1 \gg \gamma_2$, the states are separated and almost completely Gaussian. These plots show that using this specially-engineered dissipation can not only be used (along with parity measurements) to stabilise cat states, but that increasing its strength also reduces the distortions caused by the cavity self-Kerr, increasing the fidelity of the stored state. This also enables weaker pumping and smaller cats to be used without fear of the two parts of the cat overlapping, reducing exposure to other loss mechanisms.

\section{Conclusions}
We have used and extended solutions of the FPE in the generalised $P$-representation to study various system that are relevant to state of the art circuit QED experiments, with the analytical nature of the solutions allowing us to wide areas of parameter space and multiple different regimes. We have shown that a two mode cavity-transmon system can be analysed using the FPE following an adiabatic elimination of the cavity and that this method produces results that agree with other experimental and theoretical work in both the resonant and dispersive regimes, achieving good results for the steady state of the transmon and cavity even when there is strong hybridisation between the two systems. By returning to a known solution of the parametrically driven Duffing oscillator, we have studied the nature of the steady states of the system near classical phases boundaries by deriving analytical $Q$-functions. We also investigated the applications of this solution to the problems of generating squeezing in a non-ideal parametric amplifier and increasing the fidelity of Schr\"odinger cat state stabilisation. We believe that this demonstrates the potential benefits of revisiting these analytical methods as new circuit technology allows us to explore different parameter regimes, even as systems become more complex and include multiple oscillators. 

\begin{acknowledgements}
E.G. acknowledges financial support from EPSRC (EP/L026082/1).
\end{acknowledgements}

\appendix
\section{Adiabatic elimination of the cavity}
\label{app:elim}
A Fokker-Planck equation of the form
\begin{equation}
\frac{\partial P(\mathbf{x})}{\partial t} = \biggl[-\pd{x_j} A_i + \frac12 \pd{x_i}\pd{x_j} D_{ij} \biggr] P(\mathbf{x}),
\end{equation}
where $A_i$ is known as the drift vector and $D_{ij}$ the diffusion matrix, can be written equivalently as a quantum Langevin equation, provided that the diffusion matrix can be written as $D=BB^T$ for some matrix $B_{ij}$, given by

\begin{equation}
\frac{d x_i}{dt} = A_i + B_{ij} \pmb{\eta}(t),
\end{equation}
where $\pmb\eta(t)$ is a vector of zero-mean, delta correlated stochastic processes representing noise acting on the phase space coordinates. As in our system the cavity has no diffusive processes acting directly on it, the Langevin equation for this sub-system is
\begin{equation}
\pd{t} \left( \begin{array}{c}
\alpha_1\\
\beta_1 \end{array}\right)  =\left( \begin{array}{c}
-i\Delta_c \alpha_1 + \epsilon -g\alpha_2 - \frac{\gamma_c}2 \alpha_1 \\
i\Delta_c \beta_1 + \epsilon^* - g\beta_2 - \frac{\gamma_c}2 \beta_1 \end{array}  \right).
\end{equation}

In the limit that the cavity is much faster than the qubit ($\gamma_c\gg \gamma_t$) we can then assume that the cavity state relaxes extremely quickly in response to changes in the qubit field and is therefore in a steady state. By setting the equation equal to 0, we can obtain expressions for the variables of the first cavity in terms of those of the second:
\begin{equation}
\alpha_1 = \frac2{\tilde{\gamma}_c}(\epsilon - g \alpha_2) ~~~~ \beta_1 = \frac2{\tilde{\gamma}_c^*}(\epsilon^* - g \beta_2).
\end{equation}
We can use these expressions to eliminate the first mode from the system completely, by substituting them back into the FPE

\section{Solving the FPE}
\label{app:moments}
The FPE we wish to solve, after the cavity has been eliminated, is simply that for a driven, damped quantum Duffing oscillator with the parameters replaced by functions of the original system parameters. The solution is very similar to that given in \cite{Walls2008}, which we follow, but the elimination means that all of the parameters are complex and some of the simplifying solutions are not possible. The system satisfies the potential conditions $\partial F_i/\partial x_j=\partial F_j/\partial x_i$, where
\begin{equation}
F_i \equiv 2D_{ij}^{-1} \left(A_k - \frac12 \frac{\partial D_{jk}}{\partial x_k} \right),
\end{equation}
and it is therefore possible to find the steady state of the system. In this case there are only two terms to calculate,

\begin{equation}
F_1 =\frac{2}{i\chi}\left(i \chi \beta_2 +\frac{\tilde{\gamma}_t-2i\chi}{2 \alpha_2} -\frac{\tilde{\epsilon}}{\alpha_2^2}\right),
\end{equation}

\begin{equation}
F_2  =-\frac{2}{i\chi}\left(-i \chi \alpha_2 +\frac{\tilde{\gamma}^*_t + 2i\chi}{2 \beta_2} -\frac{\tilde{\epsilon}^*}{\beta_2^2}\right),
\end{equation}
and the cross derivatives are indeed equal. The steady state $P$-function is then obtained by integrating 

\begin{multline}
P_1(\pmb{\alpha})=\mathcal{N} \exp\biggl[\int\frac{1}{i\chi}\left(i \chi \beta_2 +\frac{\tilde{\gamma}_t-2i\chi}{2 \alpha_2} -\frac{\tilde{\epsilon}}{\alpha_2^2}\right) \dif \alpha_2 -
\\
\int\frac{1}{i\chi}\left(-i \chi \alpha_2 +\frac{\tilde{\gamma}^*_t + 2i\chi}{2 \beta_2} -\frac{\tilde{\epsilon}^*}{\beta_2^2}\right) \dif \beta_2\biggr]
\\
=\mathcal{N} \exp\biggl[\alpha_2\beta_2 +\left(\frac{\tilde{\gamma}_t}{2i\chi}-2\right)\log{\alpha_2} +\frac{\tilde{\epsilon}}{i\chi\alpha_2} + \alpha_2 \beta_2
\\
+\left(\frac{\tilde{\gamma}_t}{-2i\chi} -2\right)\log \beta_2 + \frac{\tilde{\epsilon}}{-i\chi\beta_2}\biggr]
\\
=\mathcal{N} \alpha_2^{d-2} \beta_2^{d^*-2} \exp{\left[\frac{\tilde{\epsilon}}{i\chi}\frac1{\alpha_2} + \frac{\tilde{\epsilon}^*}{-i\chi}\frac1{\beta_2} +2\alpha_2\beta_2\right]},
\end{multline}
where we have defined $d=\tilde{\gamma}_t/(2i\chi)$ and $\mathcal{N}$ is some normalisation constant. In order to find $\mathcal{N}$, we integrate $P({\pmb{\alpha}})$ again, making use of the substitution $x=1/\alpha_2, y = 1/\beta_2$ and Taylor expanding the second term of the exponential to give
\begin{equation}
\frac1{\mathcal{N}} = \int \sum_{n=0}^\infty \frac{2^n}{n!} x^{-d-n} y^{-d^*-n} \exp{\left[\frac{\tilde{\epsilon}}{i\chi}x + \frac{\tilde{\epsilon}^*}{-i\chi}y \right]} \dif x \dif y.
\end{equation}

These integrals are related to the Gamma function by the identity
\begin{equation}
2\pi i \frac{t^{n+d-1}}{\Gamma(d+n)} = \int_C x^{-n-d} e^{xt} \dif x,
\end{equation}
which implies that
\begin{equation}
\frac1{\mathcal{N}} = -\sum_{n=0}^\infty
\frac{2^n}{n!} \left(\frac{\tilde{\epsilon}}{i\chi}\right)^{d+n-1} \left(\frac{\tilde{\epsilon}^*}{-i\chi}\right)^{d^*+n-1} \frac{4\pi^2}{\left|\Gamma(d+n)\right|^2}, 
\end{equation}

where we have also used the fact that $\left(x^y\right)^*={x^*}^{y^*}$. Finally we note that the infinite sum is of the same form as the definition of the hypergeometric function $\tensor[_0]{F}{_2}(a,b;x)$ and that the normalisation can be written

\begin{equation}
 \frac1{\mathcal{N}} 
=-4\pi^2\left(\frac{\tilde{\epsilon}}{i\chi}\right)^{d-1}  \left(\frac{\tilde{\epsilon}^*}{-i\chi}\right)^{d^*-1}\frac{ F\left(d,d^*,2\left|\frac{\tilde{\epsilon}}{\chi}\right|^2\right) }{\Gamma(d)\Gamma(d^*)}  .
\end{equation}

The moments of the transmon field in the generalised $P$-representation are defined as
\begin{equation}
\expect{b^{\dag n}b^m } = \int\int \alpha_2^m \beta_2^n P_1(\alpha_2,\beta_2)\dif\alpha_2 \dif \beta_2,
\end{equation}
and are of the same form as the normalisation integral but with $d \to d+m, d^* \to d^*+n$. We therefore have the final expression for the moments

\begin{multline}
\expect{b^{\dag n}b^m}=
\\
\left(\frac{\tilde{\epsilon}}{i\chi}\right)^m\left(\frac{\tilde{\epsilon}^*}{-i\chi}\right)^n\frac{\Gamma(d)\Gamma(d^*)F(d+m,d^*+n,2|\frac{\tilde{\epsilon}}{\chi}|^2)}{\Gamma(d+m)\Gamma(d^*+n)F(d,d^*,2|\frac{\tilde{\epsilon}}{\chi}|^2)}.
\end{multline}

\section{Cavity Moments}
\label{app:cavmom}
The moments of the cavity field in the generalised $P$-representation are given by
\begin{equation}
\expect{a^{\dag n}a^m } = \int \alpha_1^m \beta_1^n P(\pmb{\alpha})\dif\pmb\alpha.
\end{equation}
If we instead substitute in the relations given in Eq. (\ref{eqn:elim}), then we obtain a new expression for the moments 
\begin{multline}
\expect{a^{\dag n}a^m } = \left(\frac2{\tilde{\gamma}_c}\right)^m \left(\frac2{\tilde{\gamma}_c^*}\right)^n 
\\
\int (\epsilon - g \alpha_2)^m (\epsilon^* - g \beta_2)^n P(\pmb{\alpha})\dif\pmb\alpha.
\end{multline}
This can be expanded out for any value of $m$ and $n$ and written in terms of moments of the transmon subsystem. For example, $\expect{a} (m=1,n=0)$ is given by
\begin{multline}
\expect{a} = \frac2{\tilde{\gamma}_c} \int (\epsilon - g \alpha_2) P(\pmb{\alpha})\dif\pmb\alpha
\\
= \frac2{\tilde{\gamma}_c} \left(\epsilon  \int P(\pmb{\alpha})\dif\pmb\alpha -g \int \alpha_2 P(\pmb{\alpha})\dif\pmb\alpha \right)
\\
= \frac2{\tilde{\gamma}_c}( \epsilon -g \expect{b}).
\end{multline}
where we have used the fact that the $P$-function is normalised over phase space.

\section{Transmon $P(n)$ and $Q$-functions}
\label{app:pandq}
The photon number distribution of the transmon can be written in the generalised $P$-representation as \cite{Kheruntsyan1999}
\begin{equation}
    P(n)=\frac1{n!} \iint \alpha_2^n \beta_2^n e^{-\alpha_2 \beta_2} P(\alpha_2,\beta_2) \dif \alpha_2 \dif \beta_2.
\end{equation}
This integral is the same as that for the moments, up to the coefficients of the terms of the Taylor expansion and with $m=n$, so the number distributions is given by
\begin{equation}
P(n)= \left|\frac{\tilde{\epsilon}}{\chi}\right|^{2n}\frac{\Gamma(d)\Gamma(d^*)F(d+n,d^*+n,|\frac{\tilde{\epsilon}}{\chi}|^2)}{\Gamma(d+n)\Gamma(d^*+n)F(d,d^*,2|\frac{\tilde{\epsilon}}{\chi}|^2)}.
\end{equation}

The $Q$-function is defined by performing the trace of the density matrix over a basis of coherent states
\begin{equation}
Q(\alpha) = \frac1\pi \langle \alpha|\rho|\alpha \rangle
\end{equation}
In the generalised $P$-representation, this can be written as

\begin{multline}
Q(\alpha)=e^{-\alpha \beta}
\\
\iint  \sum_{k,l=0}^\infty \frac{(-1)^{k+l}}{k! l!}\alpha^k \beta^l  \alpha'^k \beta'^l e^{-\alpha'\beta'} P(\alpha',\beta') \dif \alpha' \dif \beta' 
\end{multline}
Again, this is just an infinite sum of the type of integrals done to calculate the moments of the field, and the $Q$-function can be written as

\begin{multline}
Q(x,y) = e^{-x^2-y^2}\sum_{k,l=0}^\infty \frac{(-1)^{k+l}}{k!l!} \left(\frac{x+iy}{\sqrt{2}}\right)^k \left(\frac{x-iy}{\sqrt{2}}\right)^l
\\
\left(\frac{\tilde{\epsilon}}{i\chi}\right)^k\left(\frac{\tilde{\epsilon}^*}{-i\chi}\right)^l\frac{\Gamma(d)\Gamma(d^*)F(d+k,d^*+l,|\frac{\tilde{\epsilon}}{\chi}|^2)}{\Gamma(d+k)\Gamma(d^*+l)F(d,d^*,2|\frac{\tilde{\epsilon}}{\chi}|^2)}
\end{multline}
where, for the purposes of plotting the functions, we have written $\alpha= x+iy$.

\section{Validity of the Duffing model}
\label{app:model}
\begin{figure}[h!!!]
    \centering
    \includegraphics[width=0.49\columnwidth]{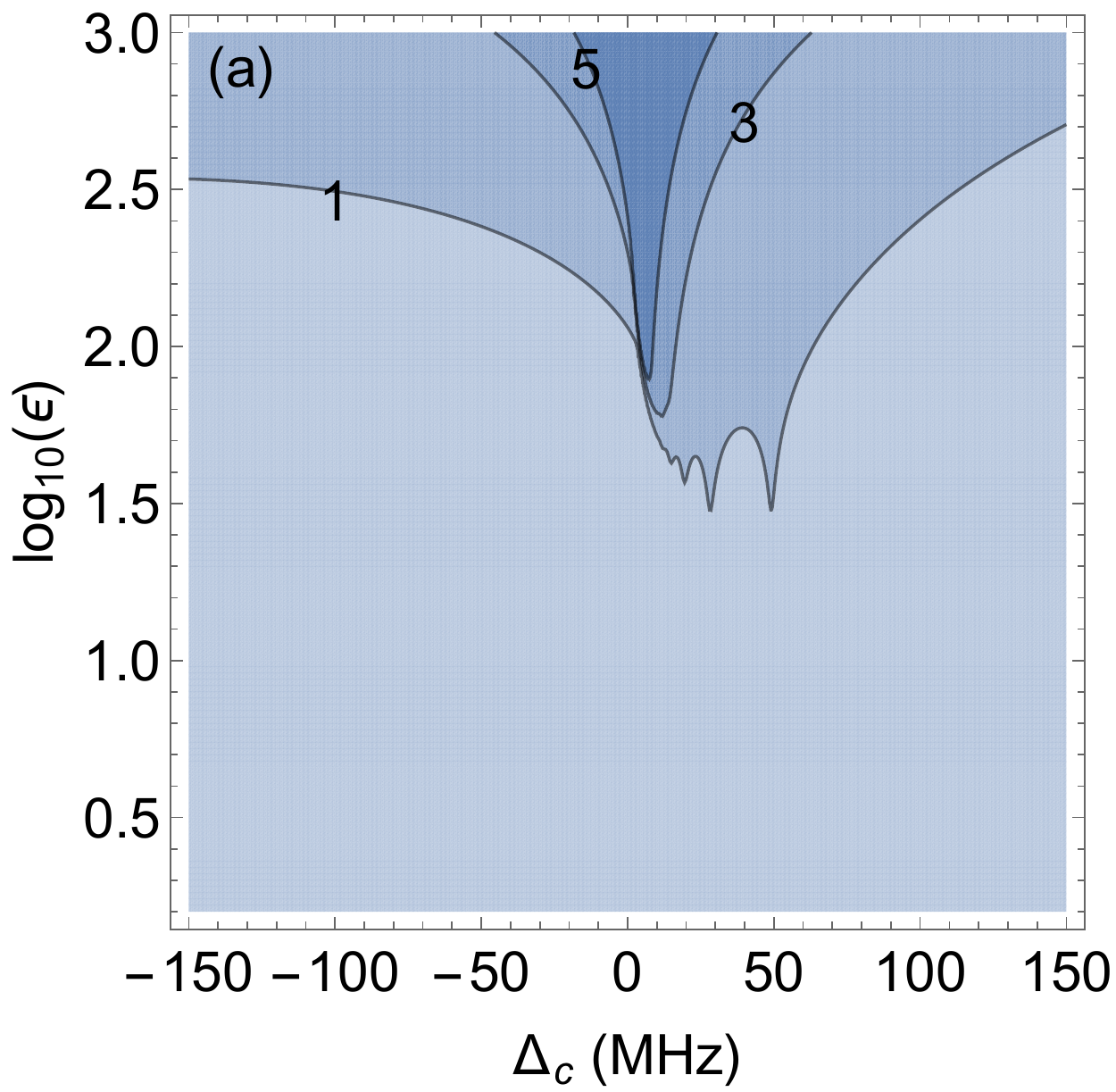}
    \includegraphics[width=0.49\columnwidth]{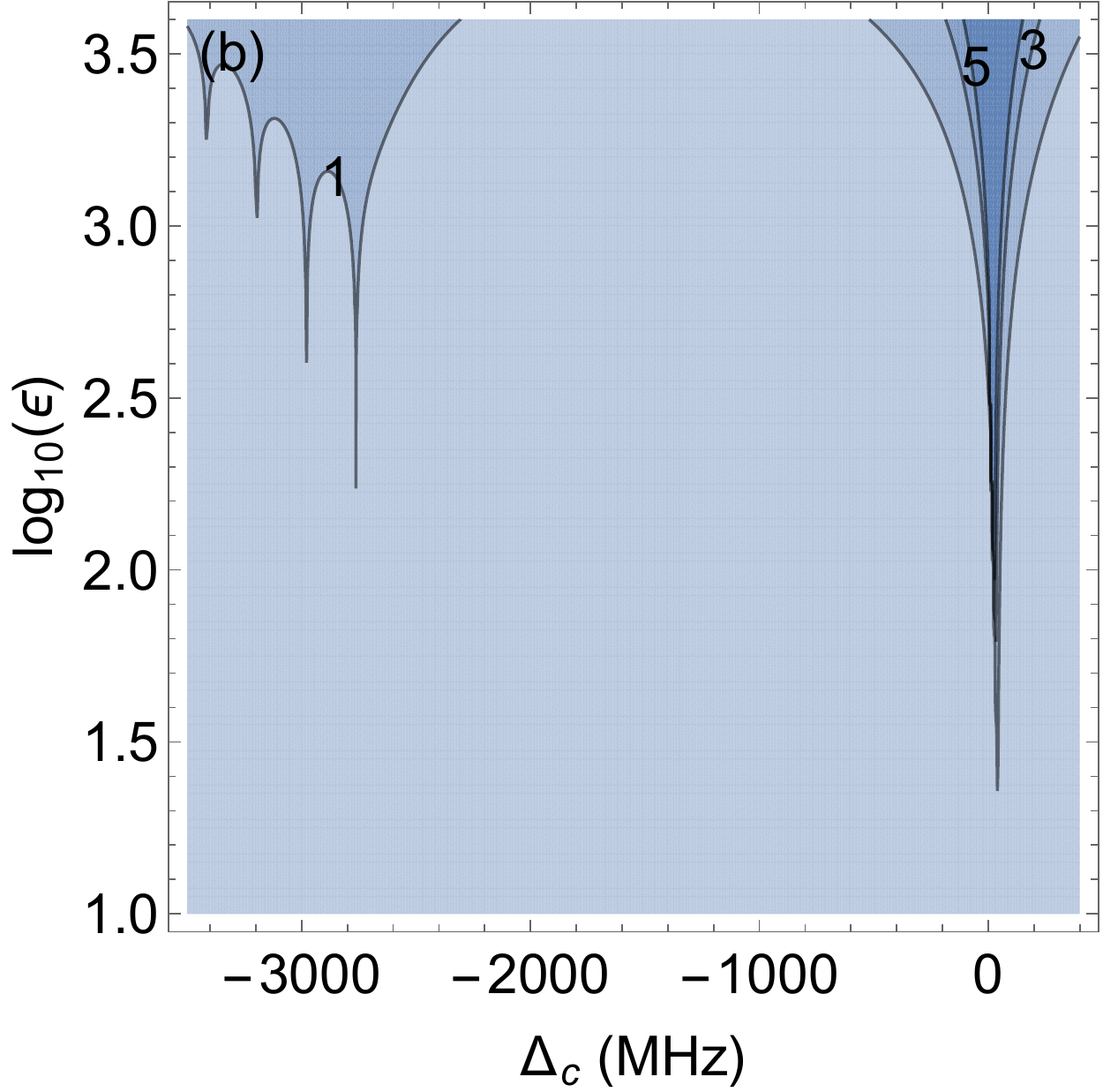}
    \caption{Plots of the number of excitations in the transmon field $\expect{b^\dag b}$ as a function of the detuning of the drive from the bare cavity $\Delta_c$ and drive amplitude $\epsilon$ in (a) the resonant regime and (b) the dispersive regime. The system parameters are the same as in Figs. \ref{fig:resonant} \& \ref{fig:dispersive1}. Contours mark boundaries with one, three and five excitations. There are fewer than three excitations over most of the parameter space, suggesting that the Duffing approximation of the transmon holds in these regions. Only the very bright peaks where the cavity frequency has almost returned to the bare cavity frequency do we see the number of excitations increase above five, suggesting that further terms in the cosine potential are required to describe these regions.}
    \label{fig:model}
\end{figure}

As discussed in Section \ref{sec:spectra}, we do not expect the Duffing model of the transmon that we use to hold for all levels of the transmon, as the higher order terms in the expansion of the cosine potential will begin to contribute significantly. If we reach a steady state of the driven dissipative system, however, where $\expect{b^\dag b}$ is kept low then these levels remain unpopulated and the accuracy of the steady state is expected to be high. Therefore, in Fig. \ref{fig:model}, we plot the number of excitations in the transmon mode for the same parameters as in Figs. \ref{fig:resonant} \& \ref{fig:dispersive1}. We see that, even while there are tens or hundreds of photons in the cavity, there are very few excitations in the transmon mode across the majority of the parameter space. As the fundamental frequency of the transmon is given by $\omega_t=\sqrt{8E_JE_C}$ \cite{Koch2007}, a good estimate of how many excited states will fit within the cosine potential, and therefore which levels are well-approximated by the Duffing model is $E_J/\omega_t=\sqrt{E_J/8E_C}$. The model therefore improves as $E_J/E_C$ is increased. In our system with $E_c=|\chi|=\SI{220}{\mega\hertz}$ and, taking $\omega_r =\SI{9.2}{\giga\hertz}$, we expect the first five excited states to be contained within the cosine potential.

In the resonant regime almost all of the features in Fig. \ref{fig:resonant}, including the supersplitting of the Rabi peaks and the movement of the transmission peak back towards the bare cavity resonance as the number of excitations increases, occur with $\expect{b^\dag b} <3$. Only when the peak has returned to within $\SI{10}{\mega\hertz}$ does the average number of excitations increase above three and excited states above the fifth begin to become significantly populated. As the power is increased further and the cavity peak becomes very bright, the average number increases greatly and the model breaks down, requiring further terms from the potential.

In the dispersive case (as shown in Figs. \ref{fig:dispersive1} and \ref{fig:dispersive2}), the peaks associated with the bare transmon transitions, along with the cavity peaks associated with the higher transmon levels, occur at low transmon occupation. The transmon bistability of Fig. \ref{fig:dip1} is also found in the region of parameter space where we expect the model to hold. The cavity resonance has shifted halfway back to the bare cavity frequency before the higher transmon levels become significantly populated. At even higher powers near this bright cavity transmission peak, higher order terms from the cosine potential should be added to the model, but the behaviour of the system still qualitatively matches experimental results from these devices.

\section{Paramp $P(n)$ and $Q$-functions}
In a very similar fashion to the cavity-transmon system above, we can also obtain an expression for the Fock state distribution of the parametrically driven Duffing oscillator
\begin{equation}
P(n) =  \frac{1}{I_{00}}\sum_{j=0}^{\infty}\frac{1}{n!j!}\left|\frac{\epsilon_2}{\kappa_2}\right|^{j+n}  
\left|\tensor[_2]{F}{_1}(-j-n,A-B,2A;2) \right|^2,
\end{equation}

where the normalisation is as defined in Eq. (\ref{eqn:ampmoms}):
\begin{equation}
I_{00}=  \sum_{j=0}^{\infty}\frac{2^j}{j!}\left|\frac{\epsilon_2}{\kappa_2}\right|^j \left|\tensor[_2]{F}{_1}(-j,A-B,2A;2)\right|^2.
\end{equation}

An analytical expression for the $Q$-function can also be written for this system and is given by 

\begin{multline}
Q(x,y) = e^{-\frac{x^2+y^2}{2}}\frac{1}{I_{00}}\sum_{j,k,l=0}^\infty \frac{(-1)^{k+l}}{k!l!} \left(\frac{x+iy}{\sqrt{2}}\right)^k 
\\
\left(\frac{x-iy}{\sqrt{2}}\right)^l \left(-\sqrt{\frac{\epsilon_2}{\kappa_2}}\right)^{j+k} \left(-\sqrt{\frac{\epsilon_2^*} {\kappa_2^*}}\right)^{j+l}
\\\tensor[_2]{F}{_1}(-j-k,A-B,2A;2)\tensor[_2]{F}{_1}(-j-l,A^*-B^*,2A^*;2).
\end{multline}

\bibliography{library}

\begin{thebibliography}{72}%
\makeatletter
\providecommand \@ifxundefined [1]{%
 \@ifx{#1\undefined}
}%
\providecommand \@ifnum [1]{%
 \ifnum #1\expandafter \@firstoftwo
 \else \expandafter \@secondoftwo
 \fi
}%
\providecommand \@ifx [1]{%
 \ifx #1\expandafter \@firstoftwo
 \else \expandafter \@secondoftwo
 \fi
}%
\providecommand \natexlab [1]{#1}%
\providecommand \enquote  [1]{``#1''}%
\providecommand \bibnamefont  [1]{#1}%
\providecommand \bibfnamefont [1]{#1}%
\providecommand \citenamefont [1]{#1}%
\providecommand \href@noop [0]{\@secondoftwo}%
\providecommand \href [0]{\begingroup \@sanitize@url \@href}%
\providecommand \@href[1]{\@@startlink{#1}\@@href}%
\providecommand \@@href[1]{\endgroup#1\@@endlink}%
\providecommand \@sanitize@url [0]{\catcode `\\12\catcode `\$12\catcode
  `\&12\catcode `\#12\catcode `\^12\catcode `\_12\catcode `\%12\relax}%
\providecommand \@@startlink[1]{}%
\providecommand \@@endlink[0]{}%
\providecommand \url  [0]{\begingroup\@sanitize@url \@url }%
\providecommand \@url [1]{\endgroup\@href {#1}{\urlprefix }}%
\providecommand \urlprefix  [0]{URL }%
\providecommand \Eprint [0]{\href }%
\providecommand \doibase [0]{http://dx.doi.org/}%
\providecommand \selectlanguage [0]{\@gobble}%
\providecommand \bibinfo  [0]{\@secondoftwo}%
\providecommand \bibfield  [0]{\@secondoftwo}%
\providecommand \translation [1]{[#1]}%
\providecommand \BibitemOpen [0]{}%
\providecommand \bibitemStop [0]{}%
\providecommand \bibitemNoStop [0]{.\EOS\space}%
\providecommand \EOS [0]{\spacefactor3000\relax}%
\providecommand \BibitemShut  [1]{\csname bibitem#1\endcsname}%
\let\auto@bib@innerbib\@empty
\bibitem [{\citenamefont {Milburn}\ and\ \citenamefont
  {Walls}(1981)}]{Milburn1981}%
  \BibitemOpen
  \bibfield  {author} {\bibinfo {author} {\bibfnamefont {G.~J.}\ \bibnamefont
  {Milburn}}\ and\ \bibinfo {author} {\bibfnamefont {D.}~\bibnamefont
  {Walls}},\ }\href@noop {} {\bibfield  {journal} {\bibinfo  {journal} {Opt.
  Comm.}\ }\textbf {\bibinfo {volume} {39}},\ \bibinfo {pages} {401} (\bibinfo
  {year} {1981})}\BibitemShut {NoStop}%
\bibitem [{\citenamefont {Drummond}\ \emph {et~al.}(1981)\citenamefont
  {Drummond}, \citenamefont {McNeil},\ and\ \citenamefont
  {Walls}}]{Drummond1981}%
  \BibitemOpen
  \bibfield  {author} {\bibinfo {author} {\bibfnamefont {P.}~\bibnamefont
  {Drummond}}, \bibinfo {author} {\bibfnamefont {K.}~\bibnamefont {McNeil}}, \
  and\ \bibinfo {author} {\bibfnamefont {D.}~\bibnamefont {Walls}},\ }\href
  {\doibase 10.1080/713820531} {\bibfield  {journal} {\bibinfo  {journal} {Opt.
  Acta Int. J. Opt.}\ }\textbf {\bibinfo {volume} {28}},\ \bibinfo {pages}
  {211} (\bibinfo {year} {1981})}\BibitemShut {NoStop}%
\bibitem [{\citenamefont {Drummond}\ and\ \citenamefont
  {Walls}(1980)}]{Drummond1980}%
  \BibitemOpen
  \bibfield  {author} {\bibinfo {author} {\bibfnamefont {P.~D.}\ \bibnamefont
  {Drummond}}\ and\ \bibinfo {author} {\bibfnamefont {D.~F.}\ \bibnamefont
  {Walls}},\ }\href {\doibase 10.1088/0305-4470/13/2/034} {\bibfield  {journal}
  {\bibinfo  {journal} {J. Phys. A. Math. Gen.}\ }\textbf {\bibinfo {volume}
  {13}},\ \bibinfo {pages} {725} (\bibinfo {year} {1980})}\BibitemShut
  {NoStop}%
\bibitem [{\citenamefont {Walls}\ and\ \citenamefont
  {Milburn}(2008)}]{Walls2008}%
  \BibitemOpen
  \bibfield  {author} {\bibinfo {author} {\bibfnamefont {D.}~\bibnamefont
  {Walls}}\ and\ \bibinfo {author} {\bibfnamefont {G.~J.}\ \bibnamefont
  {Milburn}},\ }\href@noop {} {\emph {\bibinfo {title} {{Quantum Optics}}}}\
  (\bibinfo  {publisher} {Springer-Verlag},\ \bibinfo {year}
  {2008})\BibitemShut {NoStop}%
\bibitem [{\citenamefont {Blais}\ \emph {et~al.}(2004)\citenamefont {Blais},
  \citenamefont {Huang}, \citenamefont {Wallraff}, \citenamefont {Girvin},\
  and\ \citenamefont {Schoelkopf}}]{Blais2004}%
  \BibitemOpen
  \bibfield  {author} {\bibinfo {author} {\bibfnamefont {A.}~\bibnamefont
  {Blais}}, \bibinfo {author} {\bibfnamefont {R.-S.}\ \bibnamefont {Huang}},
  \bibinfo {author} {\bibfnamefont {A.}~\bibnamefont {Wallraff}}, \bibinfo
  {author} {\bibfnamefont {S.~M.}\ \bibnamefont {Girvin}}, \ and\ \bibinfo
  {author} {\bibfnamefont {R.~J.}\ \bibnamefont {Schoelkopf}},\ }\href@noop {}
  {\bibfield  {journal} {\bibinfo  {journal} {Phys. Rev. A}\ }\textbf {\bibinfo
  {volume} {69}},\ \bibinfo {pages} {062320} (\bibinfo {year}
  {2004})}\BibitemShut {NoStop}%
\bibitem [{\citenamefont {Koch}\ \emph {et~al.}(2007)\citenamefont {Koch},
  \citenamefont {Yu}, \citenamefont {Gambetta}, \citenamefont {Houck},
  \citenamefont {Schuster}, \citenamefont {Majer}, \citenamefont {Blais},
  \citenamefont {Devoret}, \citenamefont {Girvin},\ and\ \citenamefont
  {Schoelkopf}}]{Koch2007}%
  \BibitemOpen
  \bibfield  {author} {\bibinfo {author} {\bibfnamefont {J.}~\bibnamefont
  {Koch}}, \bibinfo {author} {\bibfnamefont {T.~M.}\ \bibnamefont {Yu}},
  \bibinfo {author} {\bibfnamefont {J.}~\bibnamefont {Gambetta}}, \bibinfo
  {author} {\bibfnamefont {A.~A.}\ \bibnamefont {Houck}}, \bibinfo {author}
  {\bibfnamefont {D.~I.}\ \bibnamefont {Schuster}}, \bibinfo {author}
  {\bibfnamefont {J.}~\bibnamefont {Majer}}, \bibinfo {author} {\bibfnamefont
  {A.}~\bibnamefont {Blais}}, \bibinfo {author} {\bibfnamefont {M.~H.}\
  \bibnamefont {Devoret}}, \bibinfo {author} {\bibfnamefont {S.~M.}\
  \bibnamefont {Girvin}}, \ and\ \bibinfo {author} {\bibfnamefont {R.~J.}\
  \bibnamefont {Schoelkopf}},\ }\href@noop {} {\bibfield  {journal} {\bibinfo
  {journal} {Phys. Rev. A}\ }\textbf {\bibinfo {volume} {76}},\ \bibinfo
  {pages} {042319} (\bibinfo {year} {2007})}\BibitemShut {NoStop}%
\bibitem [{\citenamefont {Castellanos-Beltran}\ \emph
  {et~al.}(2008)\citenamefont {Castellanos-Beltran}, \citenamefont {Irwin},
  \citenamefont {Hilton}, \citenamefont {Vale},\ and\ \citenamefont
  {Lehnert}}]{Castellanos-Beltran2008}%
  \BibitemOpen
  \bibfield  {author} {\bibinfo {author} {\bibfnamefont {M.~A.}\ \bibnamefont
  {Castellanos-Beltran}}, \bibinfo {author} {\bibfnamefont {K.~D.}\
  \bibnamefont {Irwin}}, \bibinfo {author} {\bibfnamefont {G.~C.}\ \bibnamefont
  {Hilton}}, \bibinfo {author} {\bibfnamefont {L.~R.}\ \bibnamefont {Vale}}, \
  and\ \bibinfo {author} {\bibfnamefont {K.~W.}\ \bibnamefont {Lehnert}},\
  }\href {\doibase 10.1038/nphys1090} {\bibfield  {journal} {\bibinfo
  {journal} {Nat. Phys.}\ }\textbf {\bibinfo {volume} {4}},\ \bibinfo {pages}
  {13} (\bibinfo {year} {2008})}\BibitemShut {NoStop}%
\bibitem [{\citenamefont {Murch}\ \emph {et~al.}(2013)\citenamefont {Murch},
  \citenamefont {Weber}, \citenamefont {Beck}, \citenamefont {Ginossar},\ and\
  \citenamefont {Siddiqi}}]{Murch2013}%
  \BibitemOpen
  \bibfield  {author} {\bibinfo {author} {\bibfnamefont {K.~W.}\ \bibnamefont
  {Murch}}, \bibinfo {author} {\bibfnamefont {S.~J.}\ \bibnamefont {Weber}},
  \bibinfo {author} {\bibfnamefont {K.~M.}\ \bibnamefont {Beck}}, \bibinfo
  {author} {\bibfnamefont {E.}~\bibnamefont {Ginossar}}, \ and\ \bibinfo
  {author} {\bibfnamefont {I.}~\bibnamefont {Siddiqi}},\ }\href@noop {}
  {\bibfield  {journal} {\bibinfo  {journal} {Nature}\ }\textbf {\bibinfo
  {volume} {499}},\ \bibinfo {pages} {62} (\bibinfo {year} {2013})}\BibitemShut
  {NoStop}%
\bibitem [{\citenamefont {Schuster}\ \emph {et~al.}(2007)\citenamefont
  {Schuster}, \citenamefont {Houck}, \citenamefont {Schreier}, \citenamefont
  {Wallraff}, \citenamefont {Gambetta}, \citenamefont {Blais}, \citenamefont
  {Frunzio}, \citenamefont {Majer}, \citenamefont {Johnson}, \citenamefont
  {Devoret}, \citenamefont {Girvin},\ and\ \citenamefont
  {Schoelkopf}}]{Schuster2007}%
  \BibitemOpen
  \bibfield  {author} {\bibinfo {author} {\bibfnamefont {D.~I.}\ \bibnamefont
  {Schuster}}, \bibinfo {author} {\bibfnamefont {A.~A.}\ \bibnamefont {Houck}},
  \bibinfo {author} {\bibfnamefont {J.~A.}\ \bibnamefont {Schreier}}, \bibinfo
  {author} {\bibfnamefont {A.}~\bibnamefont {Wallraff}}, \bibinfo {author}
  {\bibfnamefont {J.~M.}\ \bibnamefont {Gambetta}}, \bibinfo {author}
  {\bibfnamefont {A.}~\bibnamefont {Blais}}, \bibinfo {author} {\bibfnamefont
  {L.}~\bibnamefont {Frunzio}}, \bibinfo {author} {\bibfnamefont
  {J.}~\bibnamefont {Majer}}, \bibinfo {author} {\bibfnamefont {B.~R.}\
  \bibnamefont {Johnson}}, \bibinfo {author} {\bibfnamefont {M.}~\bibnamefont
  {Devoret}}, \bibinfo {author} {\bibfnamefont {S.~M.}\ \bibnamefont {Girvin}},
  \ and\ \bibinfo {author} {\bibfnamefont {R.~J.}\ \bibnamefont {Schoelkopf}},\
  }\href@noop {} {\bibfield  {journal} {\bibinfo  {journal} {Nature}\ }\textbf
  {\bibinfo {volume} {445}},\ \bibinfo {pages} {515} (\bibinfo {year}
  {2007})}\BibitemShut {NoStop}%
\bibitem [{\citenamefont {Gambetta}\ \emph {et~al.}(2006)\citenamefont
  {Gambetta}, \citenamefont {Blais}, \citenamefont {Schuster}, \citenamefont
  {Wallraff}, \citenamefont {Frunzio}, \citenamefont {Majer}, \citenamefont
  {Devoret}, \citenamefont {Girvin},\ and\ \citenamefont
  {Schoelkopf}}]{Gambetta2006a}%
  \BibitemOpen
  \bibfield  {author} {\bibinfo {author} {\bibfnamefont {J.}~\bibnamefont
  {Gambetta}}, \bibinfo {author} {\bibfnamefont {A.}~\bibnamefont {Blais}},
  \bibinfo {author} {\bibfnamefont {D.~I.}\ \bibnamefont {Schuster}}, \bibinfo
  {author} {\bibfnamefont {A.}~\bibnamefont {Wallraff}}, \bibinfo {author}
  {\bibfnamefont {L.}~\bibnamefont {Frunzio}}, \bibinfo {author} {\bibfnamefont
  {J.}~\bibnamefont {Majer}}, \bibinfo {author} {\bibfnamefont {M.~H.}\
  \bibnamefont {Devoret}}, \bibinfo {author} {\bibfnamefont {S.~M.}\
  \bibnamefont {Girvin}}, \ and\ \bibinfo {author} {\bibfnamefont {R.~J.}\
  \bibnamefont {Schoelkopf}},\ }\href {\doibase 10.1103/PhysRevA.74.042318}
  {\bibfield  {journal} {\bibinfo  {journal} {Phys. Rev. A}\ }\textbf {\bibinfo
  {volume} {74}},\ \bibinfo {pages} {042318} (\bibinfo {year}
  {2006})}\BibitemShut {NoStop}%
\bibitem [{\citenamefont {Lutterbach}\ and\ \citenamefont
  {Davidovich}(1997)}]{Lutterbach1997}%
  \BibitemOpen
  \bibfield  {author} {\bibinfo {author} {\bibfnamefont {L.~G.}\ \bibnamefont
  {Lutterbach}}\ and\ \bibinfo {author} {\bibfnamefont {L.}~\bibnamefont
  {Davidovich}},\ }\href {\doibase 10.1103/PhysRevLett.78.2547} {\bibfield
  {journal} {\bibinfo  {journal} {Phys. Rev. Lett}\ }\textbf {\bibinfo {volume}
  {78}},\ \bibinfo {pages} {2547} (\bibinfo {year} {1997})}\BibitemShut
  {NoStop}%
\bibitem [{\citenamefont {Bertet}\ \emph {et~al.}(2002)\citenamefont {Bertet},
  \citenamefont {Auffeves}, \citenamefont {Maioli}, \citenamefont {Osnaghi},
  \citenamefont {Meunier}, \citenamefont {Brune}, \citenamefont {Raimond},\
  and\ \citenamefont {Haroche}}]{Bertet2002}%
  \BibitemOpen
  \bibfield  {author} {\bibinfo {author} {\bibfnamefont {P.}~\bibnamefont
  {Bertet}}, \bibinfo {author} {\bibfnamefont {A.}~\bibnamefont {Auffeves}},
  \bibinfo {author} {\bibfnamefont {P.}~\bibnamefont {Maioli}}, \bibinfo
  {author} {\bibfnamefont {S.}~\bibnamefont {Osnaghi}}, \bibinfo {author}
  {\bibfnamefont {T.}~\bibnamefont {Meunier}}, \bibinfo {author} {\bibfnamefont
  {M.}~\bibnamefont {Brune}}, \bibinfo {author} {\bibfnamefont {J.~M.}\
  \bibnamefont {Raimond}}, \ and\ \bibinfo {author} {\bibfnamefont
  {S.}~\bibnamefont {Haroche}},\ }\href {\doibase
  10.1103/PhysRevLett.89.200402} {\bibfield  {journal} {\bibinfo  {journal}
  {Phys. Rev. Lett.}\ }\textbf {\bibinfo {volume} {89}},\ \bibinfo {pages}
  {200402} (\bibinfo {year} {2002})}\BibitemShut {NoStop}%
\bibitem [{\citenamefont {Puri}\ and\ \citenamefont
  {Blais}(2016{\natexlab{a}})}]{Puri2016}%
  \BibitemOpen
  \bibfield  {author} {\bibinfo {author} {\bibfnamefont {S.}~\bibnamefont
  {Puri}}\ and\ \bibinfo {author} {\bibfnamefont {A.}~\bibnamefont {Blais}},\
  }\href {\doibase 10.1103/PhysRevLett.116.180501} {\bibfield  {journal}
  {\bibinfo  {journal} {Phys. Rev. Lett.}\ }\textbf {\bibinfo {volume} {116}},\
  \bibinfo {pages} {180501} (\bibinfo {year} {2016}{\natexlab{a}})}\BibitemShut
  {NoStop}%
\bibitem [{\citenamefont {Tai}\ \emph {et~al.}(2014)\citenamefont {Tai},
  \citenamefont {Lin},\ and\ \citenamefont {Goan}}]{Tai2014}%
  \BibitemOpen
  \bibfield  {author} {\bibinfo {author} {\bibfnamefont {J.~S.}\ \bibnamefont
  {Tai}}, \bibinfo {author} {\bibfnamefont {K.~T.}\ \bibnamefont {Lin}}, \ and\
  \bibinfo {author} {\bibfnamefont {H.~S.}\ \bibnamefont {Goan}},\ }\href
  {\doibase 10.1103/PhysRevA.89.062310} {\bibfield  {journal} {\bibinfo
  {journal} {Phys. Rev. A}\ }\textbf {\bibinfo {volume} {89}},\ \bibinfo
  {pages} {062310} (\bibinfo {year} {2014})}\BibitemShut {NoStop}%
\bibitem [{\citenamefont {LeBoite}\ \emph {et~al.}(2014)\citenamefont
  {LeBoite}, \citenamefont {Orso},\ and\ \citenamefont {Ciuti}}]{Boite2014}%
  \BibitemOpen
  \bibfield  {author} {\bibinfo {author} {\bibfnamefont {A.}~\bibnamefont
  {LeBoite}}, \bibinfo {author} {\bibfnamefont {G.}~\bibnamefont {Orso}}, \
  and\ \bibinfo {author} {\bibfnamefont {C.}~\bibnamefont {Ciuti}},\ }\href
  {\doibase 10.1103/PhysRevA.90.063821} {\bibfield  {journal} {\bibinfo
  {journal} {Phys. Rev. A}\ }\textbf {\bibinfo {volume} {90}},\ \bibinfo
  {pages} {063821} (\bibinfo {year} {2014})}\BibitemShut {NoStop}%
\bibitem [{\citenamefont {Jin}\ \emph {et~al.}(2014)\citenamefont {Jin},
  \citenamefont {Rossini}, \citenamefont {Leib}, \citenamefont {Hartmann},\
  and\ \citenamefont {Fazio}}]{Jin2014}%
  \BibitemOpen
  \bibfield  {author} {\bibinfo {author} {\bibfnamefont {J.}~\bibnamefont
  {Jin}}, \bibinfo {author} {\bibfnamefont {D.}~\bibnamefont {Rossini}},
  \bibinfo {author} {\bibfnamefont {M.}~\bibnamefont {Leib}}, \bibinfo {author}
  {\bibfnamefont {M.~J.}\ \bibnamefont {Hartmann}}, \ and\ \bibinfo {author}
  {\bibfnamefont {R.}~\bibnamefont {Fazio}},\ }\href {\doibase
  10.1103/PhysRevA.90.023827} {\ \textbf {\bibinfo {volume} {90}},\ \bibinfo
  {pages} {023827} (\bibinfo {year} {2014})}\BibitemShut {NoStop}%
\bibitem [{\citenamefont {Fitzpatrick}\ \emph {et~al.}(2016)\citenamefont
  {Fitzpatrick}, \citenamefont {Sundaresan}, \citenamefont {Li}, \citenamefont
  {Koch},\ and\ \citenamefont {Houck}}]{Fitzpatrick2016}%
  \BibitemOpen
  \bibfield  {author} {\bibinfo {author} {\bibfnamefont {M.}~\bibnamefont
  {Fitzpatrick}}, \bibinfo {author} {\bibfnamefont {N.~M.}\ \bibnamefont
  {Sundaresan}}, \bibinfo {author} {\bibfnamefont {A.~C.~Y.}\ \bibnamefont
  {Li}}, \bibinfo {author} {\bibfnamefont {J.}~\bibnamefont {Koch}}, \ and\
  \bibinfo {author} {\bibfnamefont {A.~A.}\ \bibnamefont {Houck}},\ }\href@noop
  {} {\  (\bibinfo {year} {2016})},\ \Eprint
  {http://arxiv.org/abs/arXiv:1607.06895v1} {arXiv:1607.06895v1} \BibitemShut
  {NoStop}%
\bibitem [{\citenamefont {Houck}\ \emph {et~al.}(2012)\citenamefont {Houck},
  \citenamefont {T{\"{u}}reci},\ and\ \citenamefont {Koch}}]{Houck2012}%
  \BibitemOpen
  \bibfield  {author} {\bibinfo {author} {\bibfnamefont {A.~a.}\ \bibnamefont
  {Houck}}, \bibinfo {author} {\bibfnamefont {H.~E.}\ \bibnamefont
  {T{\"{u}}reci}}, \ and\ \bibinfo {author} {\bibfnamefont {J.}~\bibnamefont
  {Koch}},\ }\href {\doibase 10.1038/nphys2251} {\bibfield  {journal} {\bibinfo
   {journal} {Nat. Phys.}\ }\textbf {\bibinfo {volume} {8}},\ \bibinfo {pages}
  {292} (\bibinfo {year} {2012})}\BibitemShut {NoStop}%
\bibitem [{\citenamefont {Barzanjeh}\ \emph {et~al.}(2014)\citenamefont
  {Barzanjeh}, \citenamefont {DiVincenzo},\ and\ \citenamefont
  {Terhal}}]{Barzanjeh2014}%
  \BibitemOpen
  \bibfield  {author} {\bibinfo {author} {\bibfnamefont {S.}~\bibnamefont
  {Barzanjeh}}, \bibinfo {author} {\bibfnamefont {D.~P.}\ \bibnamefont
  {DiVincenzo}}, \ and\ \bibinfo {author} {\bibfnamefont {B.~M.}\ \bibnamefont
  {Terhal}},\ }\href {\doibase 10.1103/PhysRevB.90.134515} {\bibfield
  {journal} {\bibinfo  {journal} {Phys. Rev. B}\ }\textbf {\bibinfo {volume}
  {90}},\ \bibinfo {pages} {134515} (\bibinfo {year} {2014})}\BibitemShut
  {NoStop}%
\bibitem [{\citenamefont {Khezri}\ \emph {et~al.}(2016)\citenamefont {Khezri},
  \citenamefont {Mlinar}, \citenamefont {Dressel},\ and\ \citenamefont
  {Korotkov}}]{Khezri2016}%
  \BibitemOpen
  \bibfield  {author} {\bibinfo {author} {\bibfnamefont {M.}~\bibnamefont
  {Khezri}}, \bibinfo {author} {\bibfnamefont {E.}~\bibnamefont {Mlinar}},
  \bibinfo {author} {\bibfnamefont {J.}~\bibnamefont {Dressel}}, \ and\
  \bibinfo {author} {\bibfnamefont {A.~N.}\ \bibnamefont {Korotkov}},\
  }\href@noop {} {\bibfield  {journal} {\bibinfo  {journal} {Phys. Rev. A}\
  }\textbf {\bibinfo {volume} {94}},\ \bibinfo {pages} {012347} (\bibinfo
  {year} {2016})}\BibitemShut {NoStop}%
\bibitem [{\citenamefont {Bowen}\ and\ \citenamefont {J}(2016)}]{Bowen2016}%
  \BibitemOpen
  \bibfield  {author} {\bibinfo {author} {\bibfnamefont {P.~B.}\ \bibnamefont
  {Bowen}}\ and\ \bibinfo {author} {\bibfnamefont {M.~G.}\ \bibnamefont {J}},\
  }\href@noop {} {\emph {\bibinfo {title} {{Quantum Optomechanics}}}}\
  (\bibinfo  {publisher} {CRC Press},\ \bibinfo {year} {2016})\BibitemShut
  {NoStop}%
\bibitem [{\citenamefont {Aspelmeyer}\ \emph {et~al.}(2014)\citenamefont
  {Aspelmeyer}, \citenamefont {Kippenberg},\ and\ \citenamefont
  {Marquardt}}]{Aspelmeyer2014}%
  \BibitemOpen
  \bibfield  {author} {\bibinfo {author} {\bibfnamefont {M.}~\bibnamefont
  {Aspelmeyer}}, \bibinfo {author} {\bibfnamefont {T.~J.}\ \bibnamefont
  {Kippenberg}}, \ and\ \bibinfo {author} {\bibfnamefont {F.}~\bibnamefont
  {Marquardt}},\ }\href {\doibase 10.1103/RevModPhys.86.1391} {\bibfield
  {journal} {\bibinfo  {journal} {Rev. Mod. Phys.}\ }\textbf {\bibinfo {volume}
  {86}},\ \bibinfo {pages} {1391} (\bibinfo {year} {2014})}\BibitemShut
  {NoStop}%
\bibitem [{\citenamefont {Yuan}\ \emph {et~al.}(2015)\citenamefont {Yuan},
  \citenamefont {Singh}, \citenamefont {Blanter},\ and\ \citenamefont
  {Steele}}]{Yuan2015}%
  \BibitemOpen
  \bibfield  {author} {\bibinfo {author} {\bibfnamefont {M.}~\bibnamefont
  {Yuan}}, \bibinfo {author} {\bibfnamefont {V.}~\bibnamefont {Singh}},
  \bibinfo {author} {\bibfnamefont {Y.~M.}\ \bibnamefont {Blanter}}, \ and\
  \bibinfo {author} {\bibfnamefont {G.~A.}\ \bibnamefont {Steele}},\ }\href
  {\doibase 10.1038/ncomms9491} {\bibfield  {journal} {\bibinfo  {journal}
  {Nat. Commun.}\ }\textbf {\bibinfo {volume} {6}},\ \bibinfo {pages} {8491}
  (\bibinfo {year} {2015})}\BibitemShut {NoStop}%
\bibitem [{\citenamefont {Peterson}\ \emph {et~al.}(2016)\citenamefont
  {Peterson}, \citenamefont {Purdy}, \citenamefont {Kampel}, \citenamefont
  {Andrews}, \citenamefont {Yu}, \citenamefont {Lehnert},\ and\ \citenamefont
  {Regal}}]{Peterson2016}%
  \BibitemOpen
  \bibfield  {author} {\bibinfo {author} {\bibfnamefont {R.~W.}\ \bibnamefont
  {Peterson}}, \bibinfo {author} {\bibfnamefont {T.~P.}\ \bibnamefont {Purdy}},
  \bibinfo {author} {\bibfnamefont {N.~S.}\ \bibnamefont {Kampel}}, \bibinfo
  {author} {\bibfnamefont {R.~W.}\ \bibnamefont {Andrews}}, \bibinfo {author}
  {\bibfnamefont {P.~L.}\ \bibnamefont {Yu}}, \bibinfo {author} {\bibfnamefont
  {K.~W.}\ \bibnamefont {Lehnert}}, \ and\ \bibinfo {author} {\bibfnamefont
  {C.~A.}\ \bibnamefont {Regal}},\ }\href {\doibase
  10.1103/PhysRevLett.116.063601} {\bibfield  {journal} {\bibinfo  {journal}
  {Phys. Rev. Lett.}\ }\textbf {\bibinfo {volume} {116}},\ \bibinfo {pages}
  {063601} (\bibinfo {year} {2016})}\BibitemShut {NoStop}%
\bibitem [{\citenamefont {Habibi}\ \emph {et~al.}(2016)\citenamefont {Habibi},
  \citenamefont {Zeuthen}, \citenamefont {Ghanaatshoar},\ and\ \citenamefont
  {Hammerer}}]{Habibi2016}%
  \BibitemOpen
  \bibfield  {author} {\bibinfo {author} {\bibfnamefont {H.}~\bibnamefont
  {Habibi}}, \bibinfo {author} {\bibfnamefont {E.}~\bibnamefont {Zeuthen}},
  \bibinfo {author} {\bibfnamefont {M.}~\bibnamefont {Ghanaatshoar}}, \ and\
  \bibinfo {author} {\bibfnamefont {K.}~\bibnamefont {Hammerer}},\ }\href
  {\doibase 10.1088/2040-8978/18/8/084004} {\bibfield  {journal} {\bibinfo
  {journal} {J. Opt.}\ }\textbf {\bibinfo {volume} {18}},\ \bibinfo {pages}
  {084004} (\bibinfo {year} {2016})}\BibitemShut {NoStop}%
\bibitem [{\citenamefont {Abdi}\ \emph {et~al.}(2016)\citenamefont {Abdi},
  \citenamefont {Degenfeld-Schonburg}, \citenamefont {Sameti}, \citenamefont
  {Navarrete-Benlloch},\ and\ \citenamefont {Hartmann}}]{Abdi2016}%
  \BibitemOpen
  \bibfield  {author} {\bibinfo {author} {\bibfnamefont {M.}~\bibnamefont
  {Abdi}}, \bibinfo {author} {\bibfnamefont {P.}~\bibnamefont
  {Degenfeld-Schonburg}}, \bibinfo {author} {\bibfnamefont {M.}~\bibnamefont
  {Sameti}}, \bibinfo {author} {\bibfnamefont {C.}~\bibnamefont
  {Navarrete-Benlloch}}, \ and\ \bibinfo {author} {\bibfnamefont {M.~J.}\
  \bibnamefont {Hartmann}},\ }\href {\doibase 10.1103/PhysRevLett.116.233604}
  {\bibfield  {journal} {\bibinfo  {journal} {Phys. Rev. Lett.}\ }\textbf
  {\bibinfo {volume} {116}},\ \bibinfo {pages} {233604} (\bibinfo {year}
  {2016})}\BibitemShut {NoStop}%
\bibitem [{\citenamefont {Clark}\ \emph {et~al.}(2016)\citenamefont {Clark},
  \citenamefont {Lecocq}, \citenamefont {Simmonds}, \citenamefont {Aumentado},\
  and\ \citenamefont {Teufel}}]{Clark2016}%
  \BibitemOpen
  \bibfield  {author} {\bibinfo {author} {\bibfnamefont {J.~B.}\ \bibnamefont
  {Clark}}, \bibinfo {author} {\bibfnamefont {F.}~\bibnamefont {Lecocq}},
  \bibinfo {author} {\bibfnamefont {R.~W.}\ \bibnamefont {Simmonds}}, \bibinfo
  {author} {\bibfnamefont {J.}~\bibnamefont {Aumentado}}, \ and\ \bibinfo
  {author} {\bibfnamefont {J.~D.}\ \bibnamefont {Teufel}},\ }\href {\doibase
  10.1038/NPHYS3701} {\bibfield  {journal} {\bibinfo  {journal} {Nat. Phys.}\
  }\textbf {\bibinfo {volume} {12}},\ \bibinfo {pages} {683} (\bibinfo {year}
  {2016})}\BibitemShut {NoStop}%
\bibitem [{\citenamefont {Aasi}\ \emph {et~al.}(2013)\citenamefont {Aasi} \emph
  {et~al.}}]{Aasi2013}%
  \BibitemOpen
  \bibfield  {author} {\bibinfo {author} {\bibfnamefont {J.}~\bibnamefont
  {Aasi}} \emph {et~al.},\ }\href@noop {} {\bibfield  {journal} {\bibinfo
  {journal} {Nat. Photonics}\ }\textbf {\bibinfo {volume} {7}},\ \bibinfo
  {pages} {613} (\bibinfo {year} {2013})}\BibitemShut {NoStop}%
\bibitem [{\citenamefont {Pooser}\ and\ \citenamefont
  {Lawrie}(2015)}]{Pooser2015}%
  \BibitemOpen
  \bibfield  {author} {\bibinfo {author} {\bibfnamefont {R.~C.}\ \bibnamefont
  {Pooser}}\ and\ \bibinfo {author} {\bibfnamefont {B.}~\bibnamefont
  {Lawrie}},\ }\href@noop {} {\bibfield  {journal} {\bibinfo  {journal}
  {Optica}\ }\textbf {\bibinfo {volume} {2}},\ \bibinfo {pages} {393} (\bibinfo
  {year} {2015})}\BibitemShut {NoStop}%
\bibitem [{\citenamefont {Andrews}\ \emph {et~al.}(2014)\citenamefont
  {Andrews}, \citenamefont {Peterson}, \citenamefont {Purdy}, \citenamefont
  {Cicak}, \citenamefont {Simmonds}, \citenamefont {Regal},\ and\ \citenamefont
  {Lehnert}}]{Andrews2014}%
  \BibitemOpen
  \bibfield  {author} {\bibinfo {author} {\bibfnamefont {R.~W.}\ \bibnamefont
  {Andrews}}, \bibinfo {author} {\bibfnamefont {R.~W.}\ \bibnamefont
  {Peterson}}, \bibinfo {author} {\bibfnamefont {T.~P.}\ \bibnamefont {Purdy}},
  \bibinfo {author} {\bibfnamefont {K.}~\bibnamefont {Cicak}}, \bibinfo
  {author} {\bibfnamefont {R.~W.}\ \bibnamefont {Simmonds}}, \bibinfo {author}
  {\bibfnamefont {C.~A.}\ \bibnamefont {Regal}}, \ and\ \bibinfo {author}
  {\bibfnamefont {K.~W.}\ \bibnamefont {Lehnert}},\ }\href@noop {} {\bibfield
  {journal} {\bibinfo  {journal} {Nat. Phys.}\ }\textbf {\bibinfo {volume}
  {10}},\ \bibinfo {pages} {321} (\bibinfo {year} {2014})}\BibitemShut
  {NoStop}%
\bibitem [{\citenamefont {Bishop}\ \emph {et~al.}(2010)\citenamefont {Bishop},
  \citenamefont {Ginossar},\ and\ \citenamefont {Girvin}}]{Bishop2010}%
  \BibitemOpen
  \bibfield  {author} {\bibinfo {author} {\bibfnamefont {L.~S.}\ \bibnamefont
  {Bishop}}, \bibinfo {author} {\bibfnamefont {E.}~\bibnamefont {Ginossar}}, \
  and\ \bibinfo {author} {\bibfnamefont {S.~M.}\ \bibnamefont {Girvin}},\
  }\href {\doibase 10.1103/PhysRevLett.105.100505} {\bibfield  {journal}
  {\bibinfo  {journal} {Phys. Rev. Lett.}\ }\textbf {\bibinfo {volume} {105}},\
  \bibinfo {pages} {100505} (\bibinfo {year} {2010})}\BibitemShut {NoStop}%
\bibitem [{\citenamefont {Rau}\ \emph {et~al.}(2004)\citenamefont {Rau},
  \citenamefont {Johansson},\ and\ \citenamefont {Shnirman}}]{Rau2004}%
  \BibitemOpen
  \bibfield  {author} {\bibinfo {author} {\bibfnamefont {I.}~\bibnamefont
  {Rau}}, \bibinfo {author} {\bibfnamefont {G.}~\bibnamefont {Johansson}}, \
  and\ \bibinfo {author} {\bibfnamefont {A.}~\bibnamefont {Shnirman}},\ }\href
  {\doibase 10.1103/PhysRevB.70.054521} {\bibfield  {journal} {\bibinfo
  {journal} {Phys. Rev. B}\ }\textbf {\bibinfo {volume} {70}},\ \bibinfo
  {pages} {054521} (\bibinfo {year} {2004})}\BibitemShut {NoStop}%
\bibitem [{\citenamefont {Fink}\ \emph {et~al.}(2010)\citenamefont {Fink},
  \citenamefont {Steffen}, \citenamefont {Studer}, \citenamefont {Bishop},
  \citenamefont {Baur}, \citenamefont {Bianchetti}, \citenamefont {Bozyigit},
  \citenamefont {Lang}, \citenamefont {Filipp}, \citenamefont {Leek},\ and\
  \citenamefont {Wallraff}}]{Fink2010}%
  \BibitemOpen
  \bibfield  {author} {\bibinfo {author} {\bibfnamefont {J.~M.}\ \bibnamefont
  {Fink}}, \bibinfo {author} {\bibfnamefont {L.}~\bibnamefont {Steffen}},
  \bibinfo {author} {\bibfnamefont {P.}~\bibnamefont {Studer}}, \bibinfo
  {author} {\bibfnamefont {L.~S.}\ \bibnamefont {Bishop}}, \bibinfo {author}
  {\bibfnamefont {M.}~\bibnamefont {Baur}}, \bibinfo {author} {\bibfnamefont
  {R.}~\bibnamefont {Bianchetti}}, \bibinfo {author} {\bibfnamefont
  {D.}~\bibnamefont {Bozyigit}}, \bibinfo {author} {\bibfnamefont
  {C.}~\bibnamefont {Lang}}, \bibinfo {author} {\bibfnamefont {S.}~\bibnamefont
  {Filipp}}, \bibinfo {author} {\bibfnamefont {P.~J.}\ \bibnamefont {Leek}}, \
  and\ \bibinfo {author} {\bibfnamefont {A.}~\bibnamefont {Wallraff}},\ }\href
  {\doibase 10.1103/PhysRevLett.105.163601} {\bibfield  {journal} {\bibinfo
  {journal} {Phys. Rev. Lett.}\ }\textbf {\bibinfo {volume} {105}},\ \bibinfo
  {pages} {163601} (\bibinfo {year} {2010})}\BibitemShut {NoStop}%
\bibitem [{\citenamefont {Reed}\ \emph {et~al.}(2010)\citenamefont {Reed},
  \citenamefont {DiCarlo}, \citenamefont {Johnson}, \citenamefont {Sun},
  \citenamefont {Schuster}, \citenamefont {Frunzio},\ and\ \citenamefont
  {Schoelkopf}}]{Reed2010}%
  \BibitemOpen
  \bibfield  {author} {\bibinfo {author} {\bibfnamefont {M.~D.}\ \bibnamefont
  {Reed}}, \bibinfo {author} {\bibfnamefont {L.}~\bibnamefont {DiCarlo}},
  \bibinfo {author} {\bibfnamefont {B.~R.}\ \bibnamefont {Johnson}}, \bibinfo
  {author} {\bibfnamefont {L.}~\bibnamefont {Sun}}, \bibinfo {author}
  {\bibfnamefont {D.~I.}\ \bibnamefont {Schuster}}, \bibinfo {author}
  {\bibfnamefont {L.}~\bibnamefont {Frunzio}}, \ and\ \bibinfo {author}
  {\bibfnamefont {R.~J.}\ \bibnamefont {Schoelkopf}},\ }\href {\doibase
  10.1103/PhysRevLett.105.173601} {\bibfield  {journal} {\bibinfo  {journal}
  {Phys. Rev. Lett.}\ }\textbf {\bibinfo {volume} {105}},\ \bibinfo {pages}
  {173601} (\bibinfo {year} {2010})}\BibitemShut {NoStop}%
\bibitem [{\citenamefont {Bishop}\ \emph {et~al.}(2008)\citenamefont {Bishop},
  \citenamefont {Chow}, \citenamefont {Koch}, \citenamefont {Houck},
  \citenamefont {Thuneberg}, \citenamefont {Girvin},\ and\ \citenamefont
  {Schoelkopf}}]{Bishop2008}%
  \BibitemOpen
  \bibfield  {author} {\bibinfo {author} {\bibfnamefont {L.~S.}\ \bibnamefont
  {Bishop}}, \bibinfo {author} {\bibfnamefont {J.~M.}\ \bibnamefont {Chow}},
  \bibinfo {author} {\bibfnamefont {J.}~\bibnamefont {Koch}}, \bibinfo {author}
  {\bibfnamefont {A.~A.}\ \bibnamefont {Houck}}, \bibinfo {author}
  {\bibfnamefont {E.}~\bibnamefont {Thuneberg}}, \bibinfo {author}
  {\bibfnamefont {S.~M.}\ \bibnamefont {Girvin}}, \ and\ \bibinfo {author}
  {\bibfnamefont {R.~J.}\ \bibnamefont {Schoelkopf}},\ }\href@noop {}
  {\bibfield  {journal} {\bibinfo  {journal} {Nat. Phys.}\ }\textbf {\bibinfo
  {volume} {5}},\ \bibinfo {pages} {105} (\bibinfo {year} {2008})}\BibitemShut
  {NoStop}%
\bibitem [{\citenamefont {Dykman}\ \emph {et~al.}(1998)\citenamefont {Dykman},
  \citenamefont {Maloney}, \citenamefont {Smelyanskiy},\ and\ \citenamefont
  {Silverstein}}]{Dykman1998}%
  \BibitemOpen
  \bibfield  {author} {\bibinfo {author} {\bibfnamefont {M.~I.}\ \bibnamefont
  {Dykman}}, \bibinfo {author} {\bibfnamefont {C.~M.}\ \bibnamefont {Maloney}},
  \bibinfo {author} {\bibfnamefont {V.~N.}\ \bibnamefont {Smelyanskiy}}, \ and\
  \bibinfo {author} {\bibfnamefont {M.}~\bibnamefont {Silverstein}},\
  }\href@noop {} {\bibfield  {journal} {\bibinfo  {journal} {Phys. Rev. E}\
  }\textbf {\bibinfo {volume} {57}},\ \bibinfo {pages} {5202} (\bibinfo {year}
  {1998})}\BibitemShut {NoStop}%
\bibitem [{\citenamefont {Zorin}\ and\ \citenamefont
  {Makhlin}(2011)}]{Zorin2011}%
  \BibitemOpen
  \bibfield  {author} {\bibinfo {author} {\bibfnamefont {A.~B.}\ \bibnamefont
  {Zorin}}\ and\ \bibinfo {author} {\bibfnamefont {Y.}~\bibnamefont
  {Makhlin}},\ }\href {\doibase 10.1103/PhysRevB.83.224506} {\bibfield
  {journal} {\bibinfo  {journal} {Phys. Rev. B}\ }\textbf {\bibinfo {volume}
  {83}},\ \bibinfo {pages} {224506} (\bibinfo {year} {2011})}\BibitemShut
  {NoStop}%
\bibitem [{\citenamefont {Krantz}\ \emph {et~al.}(2016)\citenamefont {Krantz},
  \citenamefont {Bengtsson}, \citenamefont {Simoen}, \citenamefont
  {Gustavsson}, \citenamefont {Shumeiko}, \citenamefont {Oliver}, \citenamefont
  {Wilson}, \citenamefont {Delsing},\ and\ \citenamefont
  {Bylander}}]{Krantz2016}%
  \BibitemOpen
  \bibfield  {author} {\bibinfo {author} {\bibfnamefont {P.}~\bibnamefont
  {Krantz}}, \bibinfo {author} {\bibfnamefont {A.}~\bibnamefont {Bengtsson}},
  \bibinfo {author} {\bibfnamefont {M.}~\bibnamefont {Simoen}}, \bibinfo
  {author} {\bibfnamefont {S.}~\bibnamefont {Gustavsson}}, \bibinfo {author}
  {\bibfnamefont {V.}~\bibnamefont {Shumeiko}}, \bibinfo {author}
  {\bibfnamefont {W.~D.}\ \bibnamefont {Oliver}}, \bibinfo {author}
  {\bibfnamefont {C.~M.}\ \bibnamefont {Wilson}}, \bibinfo {author}
  {\bibfnamefont {P.}~\bibnamefont {Delsing}}, \ and\ \bibinfo {author}
  {\bibfnamefont {J.}~\bibnamefont {Bylander}},\ }\href {\doibase
  10.1038/ncomms11417} {\bibfield  {journal} {\bibinfo  {journal} {Nat.
  Commun.}\ }\textbf {\bibinfo {volume} {7}},\ \bibinfo {pages} {114417}
  (\bibinfo {year} {2016})}\BibitemShut {NoStop}%
\bibitem [{\citenamefont {Leghtas}\ \emph {et~al.}(2015)\citenamefont
  {Leghtas}, \citenamefont {Touzard}, \citenamefont {Pop}, \citenamefont {Kou},
  \citenamefont {Vlastakis}, \citenamefont {Petrenko}, \citenamefont {Sliwa},
  \citenamefont {Narla}, \citenamefont {Shankar}, \citenamefont {Hatridge},
  \citenamefont {Reagor}, \citenamefont {Frunzio}, \citenamefont {Schoelkopf},
  \citenamefont {Mirrahimi},\ and\ \citenamefont {Devoret}}]{Leghtas2015}%
  \BibitemOpen
  \bibfield  {author} {\bibinfo {author} {\bibfnamefont {Z.}~\bibnamefont
  {Leghtas}}, \bibinfo {author} {\bibfnamefont {S.}~\bibnamefont {Touzard}},
  \bibinfo {author} {\bibfnamefont {I.~M.}\ \bibnamefont {Pop}}, \bibinfo
  {author} {\bibfnamefont {A.}~\bibnamefont {Kou}}, \bibinfo {author}
  {\bibfnamefont {B.}~\bibnamefont {Vlastakis}}, \bibinfo {author}
  {\bibfnamefont {A.}~\bibnamefont {Petrenko}}, \bibinfo {author}
  {\bibfnamefont {K.~M.}\ \bibnamefont {Sliwa}}, \bibinfo {author}
  {\bibfnamefont {A.}~\bibnamefont {Narla}}, \bibinfo {author} {\bibfnamefont
  {S.}~\bibnamefont {Shankar}}, \bibinfo {author} {\bibfnamefont {M.~J.}\
  \bibnamefont {Hatridge}}, \bibinfo {author} {\bibfnamefont {M.}~\bibnamefont
  {Reagor}}, \bibinfo {author} {\bibfnamefont {L.}~\bibnamefont {Frunzio}},
  \bibinfo {author} {\bibfnamefont {R.~J.}\ \bibnamefont {Schoelkopf}},
  \bibinfo {author} {\bibfnamefont {M.}~\bibnamefont {Mirrahimi}}, \ and\
  \bibinfo {author} {\bibfnamefont {M.~H.}\ \bibnamefont {Devoret}},\ }\href
  {\doibase 10.1126/science.aaa2085} {\bibfield  {journal} {\bibinfo  {journal}
  {Science}\ }\textbf {\bibinfo {volume} {347}},\ \bibinfo {pages} {853}
  (\bibinfo {year} {2015})}\BibitemShut {NoStop}%
\bibitem [{\citenamefont {Kirchmair}\ \emph {et~al.}(2013)\citenamefont
  {Kirchmair}, \citenamefont {Vlastakis}, \citenamefont {Leghtas},
  \citenamefont {Nigg}, \citenamefont {Paik}, \citenamefont {Ginossar},
  \citenamefont {Mirrahimi}, \citenamefont {Frunzio}, \citenamefont {Girvin},\
  and\ \citenamefont {Schoelkopf}}]{Kirchmair2013}%
  \BibitemOpen
  \bibfield  {author} {\bibinfo {author} {\bibfnamefont {G.}~\bibnamefont
  {Kirchmair}}, \bibinfo {author} {\bibfnamefont {B.}~\bibnamefont
  {Vlastakis}}, \bibinfo {author} {\bibfnamefont {Z.}~\bibnamefont {Leghtas}},
  \bibinfo {author} {\bibfnamefont {S.~E.}\ \bibnamefont {Nigg}}, \bibinfo
  {author} {\bibfnamefont {H.}~\bibnamefont {Paik}}, \bibinfo {author}
  {\bibfnamefont {E.}~\bibnamefont {Ginossar}}, \bibinfo {author}
  {\bibfnamefont {M.}~\bibnamefont {Mirrahimi}}, \bibinfo {author}
  {\bibfnamefont {L.}~\bibnamefont {Frunzio}}, \bibinfo {author} {\bibfnamefont
  {S.~M.}\ \bibnamefont {Girvin}}, \ and\ \bibinfo {author} {\bibfnamefont
  {R.~J.}\ \bibnamefont {Schoelkopf}},\ }\href@noop {} {\bibfield  {journal}
  {\bibinfo  {journal} {Nature}\ }\textbf {\bibinfo {volume} {495}},\ \bibinfo
  {pages} {205} (\bibinfo {year} {2013})}\BibitemShut {NoStop}%
\bibitem [{\citenamefont {Clarke}\ and\ \citenamefont
  {Wilhelm}(2008)}]{Clarke2008}%
  \BibitemOpen
  \bibfield  {author} {\bibinfo {author} {\bibfnamefont {J.}~\bibnamefont
  {Clarke}}\ and\ \bibinfo {author} {\bibfnamefont {F.~K.}\ \bibnamefont
  {Wilhelm}},\ }\href {\doibase 10.1038/nature07128} {\bibfield  {journal}
  {\bibinfo  {journal} {Nature}\ }\textbf {\bibinfo {volume} {453}},\ \bibinfo
  {pages} {1031} (\bibinfo {year} {2008})}\BibitemShut {NoStop}%
\bibitem [{\citenamefont {Paik}\ \emph {et~al.}(2011)\citenamefont {Paik},
  \citenamefont {Schuster}, \citenamefont {Bishop}, \citenamefont {Kirchmair},
  \citenamefont {Catelani}, \citenamefont {Sears}, \citenamefont {Johnson},
  \citenamefont {Reagor}, \citenamefont {Frunzio}, \citenamefont {Glazman},
  \citenamefont {Girvin}, \citenamefont {Devoret},\ and\ \citenamefont
  {Schoelkopf}}]{Paik2011}%
  \BibitemOpen
  \bibfield  {author} {\bibinfo {author} {\bibfnamefont {H.}~\bibnamefont
  {Paik}}, \bibinfo {author} {\bibfnamefont {D.~I.}\ \bibnamefont {Schuster}},
  \bibinfo {author} {\bibfnamefont {L.~S.}\ \bibnamefont {Bishop}}, \bibinfo
  {author} {\bibfnamefont {G.}~\bibnamefont {Kirchmair}}, \bibinfo {author}
  {\bibfnamefont {G.}~\bibnamefont {Catelani}}, \bibinfo {author}
  {\bibfnamefont {A.~P.}\ \bibnamefont {Sears}}, \bibinfo {author}
  {\bibfnamefont {B.~R.}\ \bibnamefont {Johnson}}, \bibinfo {author}
  {\bibfnamefont {M.~J.}\ \bibnamefont {Reagor}}, \bibinfo {author}
  {\bibfnamefont {L.}~\bibnamefont {Frunzio}}, \bibinfo {author} {\bibfnamefont
  {L.~I.}\ \bibnamefont {Glazman}}, \bibinfo {author} {\bibfnamefont {S.~M.}\
  \bibnamefont {Girvin}}, \bibinfo {author} {\bibfnamefont {M.~H.}\
  \bibnamefont {Devoret}}, \ and\ \bibinfo {author} {\bibfnamefont {R.~J.}\
  \bibnamefont {Schoelkopf}},\ }\href@noop {} {\bibfield  {journal} {\bibinfo
  {journal} {Phys. Rev. Lett.}\ }\textbf {\bibinfo {volume} {107}},\ \bibinfo
  {pages} {240501} (\bibinfo {year} {2011})}\BibitemShut {NoStop}%
\bibitem [{\citenamefont {Wang}\ \emph {et~al.}(2016)\citenamefont {Wang},
  \citenamefont {Gao}, \citenamefont {Reinhold}, \citenamefont {Heeres},
  \citenamefont {Ofek}, \citenamefont {Chou}, \citenamefont {Axline},
  \citenamefont {Reagor}, \citenamefont {Blumoff}, \citenamefont {Sliwa},
  \citenamefont {Frunzio}, \citenamefont {Girvin}, \citenamefont {Jiang},
  \citenamefont {Mirrahimi}, \citenamefont {Devoret},\ and\ \citenamefont
  {Schoelkopf}}]{Wang2016}%
  \BibitemOpen
  \bibfield  {author} {\bibinfo {author} {\bibfnamefont {C.}~\bibnamefont
  {Wang}}, \bibinfo {author} {\bibfnamefont {Y.~Y.}\ \bibnamefont {Gao}},
  \bibinfo {author} {\bibfnamefont {P.}~\bibnamefont {Reinhold}}, \bibinfo
  {author} {\bibfnamefont {R.~W.}\ \bibnamefont {Heeres}}, \bibinfo {author}
  {\bibfnamefont {N.}~\bibnamefont {Ofek}}, \bibinfo {author} {\bibfnamefont
  {K.}~\bibnamefont {Chou}}, \bibinfo {author} {\bibfnamefont {C.}~\bibnamefont
  {Axline}}, \bibinfo {author} {\bibfnamefont {M.}~\bibnamefont {Reagor}},
  \bibinfo {author} {\bibfnamefont {J.}~\bibnamefont {Blumoff}}, \bibinfo
  {author} {\bibfnamefont {K.~M.}\ \bibnamefont {Sliwa}}, \bibinfo {author}
  {\bibfnamefont {L.}~\bibnamefont {Frunzio}}, \bibinfo {author} {\bibfnamefont
  {S.~M.}\ \bibnamefont {Girvin}}, \bibinfo {author} {\bibfnamefont
  {L.}~\bibnamefont {Jiang}}, \bibinfo {author} {\bibfnamefont
  {M.}~\bibnamefont {Mirrahimi}}, \bibinfo {author} {\bibfnamefont {M.~H.}\
  \bibnamefont {Devoret}}, \ and\ \bibinfo {author} {\bibfnamefont {R.~J.}\
  \bibnamefont {Schoelkopf}},\ }\href {\doibase 10.1126/science.aaf2941}
  {\bibfield  {journal} {\bibinfo  {journal} {Science}\ }\textbf {\bibinfo
  {volume} {352}},\ \bibinfo {pages} {1087} (\bibinfo {year}
  {2016})}\BibitemShut {NoStop}%
\bibitem [{\citenamefont {Suri}\ \emph {et~al.}(2015)\citenamefont {Suri},
  \citenamefont {Keane}, \citenamefont {Bishop}, \citenamefont {Novikov},
  \citenamefont {Wellstood},\ and\ \citenamefont {Palmer}}]{Suri2015}%
  \BibitemOpen
  \bibfield  {author} {\bibinfo {author} {\bibfnamefont {B.}~\bibnamefont
  {Suri}}, \bibinfo {author} {\bibfnamefont {Z.~K.}\ \bibnamefont {Keane}},
  \bibinfo {author} {\bibfnamefont {L.~S.}\ \bibnamefont {Bishop}}, \bibinfo
  {author} {\bibfnamefont {S.}~\bibnamefont {Novikov}}, \bibinfo {author}
  {\bibfnamefont {F.~C.}\ \bibnamefont {Wellstood}}, \ and\ \bibinfo {author}
  {\bibfnamefont {B.~S.}\ \bibnamefont {Palmer}},\ }\href {\doibase
  10.1103/PhysRevA.92.063801} {\bibfield  {journal} {\bibinfo  {journal} {Phys.
  Rev. A}\ }\textbf {\bibinfo {volume} {92}},\ \bibinfo {pages} {063801}
  (\bibinfo {year} {2015})}\BibitemShut {NoStop}%
\bibitem [{\citenamefont {Riste}\ \emph {et~al.}(2015)\citenamefont {Riste},
  \citenamefont {Poletto}, \citenamefont {Bruno}, \citenamefont {Vesterinen},
  \citenamefont {Saire},\ and\ \citenamefont {Dicarlo}}]{Riste2015}%
  \BibitemOpen
  \bibfield  {author} {\bibinfo {author} {\bibfnamefont {D.}~\bibnamefont
  {Riste}}, \bibinfo {author} {\bibfnamefont {S.}~\bibnamefont {Poletto}},
  \bibinfo {author} {\bibfnamefont {A.}~\bibnamefont {Bruno}}, \bibinfo
  {author} {\bibfnamefont {V.}~\bibnamefont {Vesterinen}}, \bibinfo {author}
  {\bibfnamefont {O.-P.}\ \bibnamefont {Saire}}, \ and\ \bibinfo {author}
  {\bibfnamefont {L.}~\bibnamefont {Dicarlo}},\ }\href {\doibase
  10.1038/ncomms7983} {\bibfield  {journal} {\bibinfo  {journal} {Nat.
  Commun.}\ }\textbf {\bibinfo {volume} {6}},\ \bibinfo {pages} {6983}
  (\bibinfo {year} {2015})}\BibitemShut {NoStop}%
\bibitem [{\citenamefont {Bishop}(2010)}]{Bishop2010a}%
  \BibitemOpen
  \bibfield  {author} {\bibinfo {author} {\bibfnamefont {L.~S.}\ \bibnamefont
  {Bishop}},\ }\emph {\bibinfo {title} {{Circuit Quantum Electrodynamics}}},\
  \href@noop {} {Ph.D. thesis},\ \bibinfo  {school} {Yale University} (\bibinfo
  {year} {2010})\BibitemShut {NoStop}%
\bibitem [{\citenamefont {Drummond}\ and\ \citenamefont
  {Gardiner}(1980)}]{Drummond1980a}%
  \BibitemOpen
  \bibfield  {author} {\bibinfo {author} {\bibfnamefont {P.~D.}\ \bibnamefont
  {Drummond}}\ and\ \bibinfo {author} {\bibfnamefont {C.~W.}\ \bibnamefont
  {Gardiner}},\ }\href {\doibase 10.1088/0305-4470/13/7/018} {\bibfield
  {journal} {\bibinfo  {journal} {J. Phys. A. Math. Gen.}\ }\textbf {\bibinfo
  {volume} {13}},\ \bibinfo {pages} {2353} (\bibinfo {year}
  {1980})}\BibitemShut {NoStop}%
\bibitem [{\citenamefont {Mavrogordatos}\ \emph {et~al.}()\citenamefont
  {Mavrogordatos}, \citenamefont {Tancredi}, \citenamefont {Elliott},
  \citenamefont {Peterer}, \citenamefont {Patterson}, \citenamefont {Rahamim},
  \citenamefont {Leek}, \citenamefont {Ginossar},\ and\ \citenamefont
  {Szymanska}}]{Themis2016}%
  \BibitemOpen
  \bibfield  {author} {\bibinfo {author} {\bibfnamefont {T.~K.}\ \bibnamefont
  {Mavrogordatos}}, \bibinfo {author} {\bibfnamefont {G.}~\bibnamefont
  {Tancredi}}, \bibinfo {author} {\bibfnamefont {M.}~\bibnamefont {Elliott}},
  \bibinfo {author} {\bibfnamefont {M.~J.}\ \bibnamefont {Peterer}}, \bibinfo
  {author} {\bibfnamefont {A.}~\bibnamefont {Patterson}}, \bibinfo {author}
  {\bibfnamefont {J.}~\bibnamefont {Rahamim}}, \bibinfo {author} {\bibfnamefont
  {P.}~\bibnamefont {Leek}}, \bibinfo {author} {\bibfnamefont {E.}~\bibnamefont
  {Ginossar}}, \ and\ \bibinfo {author} {\bibfnamefont {M.~H.}\ \bibnamefont
  {Szymanska}},\ }\href@noop {} {\enquote {\bibinfo {title} {Simultaneous
  bistability of qubit and resonator in circuit quantum electrodynamics},}\
  }\bibinfo {note} {(unpublished)}\BibitemShut {NoStop}%
\bibitem [{\citenamefont {Braum\"uller}\ \emph {et~al.}(2015)\citenamefont
  {Braum\"uller}, \citenamefont {Cramer}, \citenamefont {Schl\"or},
  \citenamefont {Rotzinger}, \citenamefont {Radtke}, \citenamefont
  {Lukashenko}, \citenamefont {Yang}, \citenamefont {Skacel}, \citenamefont
  {Probst}, \citenamefont {Marthaler}, \citenamefont {Guo}, \citenamefont
  {Ustinov},\ and\ \citenamefont {Weides}}]{Braumuller2015}%
  \BibitemOpen
  \bibfield  {author} {\bibinfo {author} {\bibfnamefont {J.}~\bibnamefont
  {Braum\"uller}}, \bibinfo {author} {\bibfnamefont {J.}~\bibnamefont
  {Cramer}}, \bibinfo {author} {\bibfnamefont {S.}~\bibnamefont {Schl\"or}},
  \bibinfo {author} {\bibfnamefont {H.}~\bibnamefont {Rotzinger}}, \bibinfo
  {author} {\bibfnamefont {L.}~\bibnamefont {Radtke}}, \bibinfo {author}
  {\bibfnamefont {A.}~\bibnamefont {Lukashenko}}, \bibinfo {author}
  {\bibfnamefont {P.}~\bibnamefont {Yang}}, \bibinfo {author} {\bibfnamefont
  {S.~T.}\ \bibnamefont {Skacel}}, \bibinfo {author} {\bibfnamefont
  {S.}~\bibnamefont {Probst}}, \bibinfo {author} {\bibfnamefont
  {M.}~\bibnamefont {Marthaler}}, \bibinfo {author} {\bibfnamefont
  {L.}~\bibnamefont {Guo}}, \bibinfo {author} {\bibfnamefont {A.~V.}\
  \bibnamefont {Ustinov}}, \ and\ \bibinfo {author} {\bibfnamefont
  {M.}~\bibnamefont {Weides}},\ }\href@noop {} {\bibfield  {journal} {\bibinfo
  {journal} {Phys. Rev. B}\ }\textbf {\bibinfo {volume} {91}},\ \bibinfo
  {pages} {054523} (\bibinfo {year} {2015})}\BibitemShut {NoStop}%
\bibitem [{\citenamefont {Peterer}\ \emph {et~al.}(2015)\citenamefont
  {Peterer}, \citenamefont {Bader}, \citenamefont {Jin}, \citenamefont {Yan},
  \citenamefont {Kamal}, \citenamefont {Gudmundsen}, \citenamefont {Leek},
  \citenamefont {Orlando}, \citenamefont {Oliver},\ and\ \citenamefont
  {Gustavsson}}]{Peterer2015}%
  \BibitemOpen
  \bibfield  {author} {\bibinfo {author} {\bibfnamefont {M.~J.}\ \bibnamefont
  {Peterer}}, \bibinfo {author} {\bibfnamefont {S.~J.}\ \bibnamefont {Bader}},
  \bibinfo {author} {\bibfnamefont {X.}~\bibnamefont {Jin}}, \bibinfo {author}
  {\bibfnamefont {F.}~\bibnamefont {Yan}}, \bibinfo {author} {\bibfnamefont
  {A.}~\bibnamefont {Kamal}}, \bibinfo {author} {\bibfnamefont {T.~J.}\
  \bibnamefont {Gudmundsen}}, \bibinfo {author} {\bibfnamefont {P.~J.}\
  \bibnamefont {Leek}}, \bibinfo {author} {\bibfnamefont {T.~P.}\ \bibnamefont
  {Orlando}}, \bibinfo {author} {\bibfnamefont {W.~D.}\ \bibnamefont {Oliver}},
  \ and\ \bibinfo {author} {\bibfnamefont {S.}~\bibnamefont {Gustavsson}},\
  }\href {\doibase 10.1103/PhysRevLett.114.010501} {\bibfield  {journal}
  {\bibinfo  {journal} {Phys. Rev. Lett}\ }\textbf {\bibinfo {volume} {114}},\
  \bibinfo {pages} {010501} (\bibinfo {year} {2015})}\BibitemShut {NoStop}%
\bibitem [{\citenamefont {Lin}\ \emph {et~al.}(2015)\citenamefont {Lin},
  \citenamefont {Nakamura},\ and\ \citenamefont {Dykman}}]{Lin2015}%
  \BibitemOpen
  \bibfield  {author} {\bibinfo {author} {\bibfnamefont {Z.~R.}\ \bibnamefont
  {Lin}}, \bibinfo {author} {\bibfnamefont {Y.}~\bibnamefont {Nakamura}}, \
  and\ \bibinfo {author} {\bibfnamefont {M.~I.}\ \bibnamefont {Dykman}},\
  }\href {\doibase 10.1103/PhysRevE.92.022105} {\bibfield  {journal} {\bibinfo
  {journal} {Phys. Rev. E}\ }\textbf {\bibinfo {volume} {92}},\ \bibinfo
  {pages} {022105} (\bibinfo {year} {2015})}\BibitemShut {NoStop}%
\bibitem [{\citenamefont {Yurke}\ \emph {et~al.}(1988)\citenamefont {Yurke},
  \citenamefont {Kaminsky}, \citenamefont {Miller}, \citenamefont {Whittaker},
  \citenamefont {Smith}, \citenamefont {Silver},\ and\ \citenamefont
  {Simon}}]{Yurke1988}%
  \BibitemOpen
  \bibfield  {author} {\bibinfo {author} {\bibfnamefont {B.}~\bibnamefont
  {Yurke}}, \bibinfo {author} {\bibfnamefont {P.~G.}\ \bibnamefont {Kaminsky}},
  \bibinfo {author} {\bibfnamefont {R.~E.}\ \bibnamefont {Miller}}, \bibinfo
  {author} {\bibfnamefont {E.~A.}\ \bibnamefont {Whittaker}}, \bibinfo {author}
  {\bibfnamefont {A.~D.}\ \bibnamefont {Smith}}, \bibinfo {author}
  {\bibfnamefont {A.~H.}\ \bibnamefont {Silver}}, \ and\ \bibinfo {author}
  {\bibfnamefont {R.~W.}\ \bibnamefont {Simon}},\ }\href@noop {} {\bibfield
  {journal} {\bibinfo  {journal} {Phys. Rev. Lett.}\ }\textbf {\bibinfo
  {volume} {60}},\ \bibinfo {pages} {764} (\bibinfo {year} {1988})}\BibitemShut
  {NoStop}%
\bibitem [{\citenamefont {Lin}\ \emph {et~al.}(2014)\citenamefont {Lin},
  \citenamefont {Inomata}, \citenamefont {Koshino}, \citenamefont {Oliver},
  \citenamefont {Nakamura}, \citenamefont {Tsai},\ and\ \citenamefont
  {Yamamoto}}]{Lin2014}%
  \BibitemOpen
  \bibfield  {author} {\bibinfo {author} {\bibfnamefont {Z.~R.}\ \bibnamefont
  {Lin}}, \bibinfo {author} {\bibfnamefont {K.}~\bibnamefont {Inomata}},
  \bibinfo {author} {\bibfnamefont {K.}~\bibnamefont {Koshino}}, \bibinfo
  {author} {\bibfnamefont {W.~D.}\ \bibnamefont {Oliver}}, \bibinfo {author}
  {\bibfnamefont {Y.}~\bibnamefont {Nakamura}}, \bibinfo {author}
  {\bibfnamefont {J.~S.}\ \bibnamefont {Tsai}}, \ and\ \bibinfo {author}
  {\bibfnamefont {T.}~\bibnamefont {Yamamoto}},\ }\href {\doibase
  10.1038/ncomms5480} {\bibfield  {journal} {\bibinfo  {journal} {Nat.
  Commun.}\ }\textbf {\bibinfo {volume} {5}},\ \bibinfo {pages} {4480}
  (\bibinfo {year} {2014})}\BibitemShut {NoStop}%
\bibitem [{\citenamefont {Wustmann}\ and\ \citenamefont
  {Shumeiko}(2013)}]{Wustmann2013}%
  \BibitemOpen
  \bibfield  {author} {\bibinfo {author} {\bibfnamefont {W.}~\bibnamefont
  {Wustmann}}\ and\ \bibinfo {author} {\bibfnamefont {V.}~\bibnamefont
  {Shumeiko}},\ }\href {\doibase 10.1103/PhysRevB.87.184501} {\bibfield
  {journal} {\bibinfo  {journal} {Phyical Rev. B}\ }\textbf {\bibinfo {volume}
  {87}},\ \bibinfo {pages} {184501} (\bibinfo {year} {2013})}\BibitemShut
  {NoStop}%
\bibitem [{\citenamefont {Dykman}(2012)}]{Dykman2012}%
  \BibitemOpen
  \bibfield  {author} {\bibinfo {author} {\bibfnamefont {M.~I.}\ \bibnamefont
  {Dykman}},\ }in\ \href@noop {} {\emph {\bibinfo {booktitle} {Fluctuating
  nonlinear oscillators}}},\ \bibinfo {editor} {edited by\ \bibinfo {editor}
  {\bibfnamefont {M.~I.}\ \bibnamefont {Dykman}}}\ (\bibinfo  {publisher}
  {Oxford University Press},\ \bibinfo {year} {2012})\BibitemShut {NoStop}%
\bibitem [{\citenamefont {Dykman}(2007)}]{Dykman2007}%
  \BibitemOpen
  \bibfield  {author} {\bibinfo {author} {\bibfnamefont {M.~I.}\ \bibnamefont
  {Dykman}},\ }\href {\doibase 10.1103/PhysRevE.75.011101} {\bibfield
  {journal} {\bibinfo  {journal} {Phys. Rev. E}\ }\textbf {\bibinfo {volume}
  {75}},\ \bibinfo {pages} {011101} (\bibinfo {year} {2007})}\BibitemShut
  {NoStop}%
\bibitem [{\citenamefont {Hirsch}\ \emph {et~al.}(2004)\citenamefont {Hirsch},
  \citenamefont {Smale},\ and\ \citenamefont {Devaney}}]{Hirsch2004}%
  \BibitemOpen
  \bibfield  {author} {\bibinfo {author} {\bibfnamefont {M.~W.}\ \bibnamefont
  {Hirsch}}, \bibinfo {author} {\bibfnamefont {S.}~\bibnamefont {Smale}}, \
  and\ \bibinfo {author} {\bibfnamefont {R.~L.}\ \bibnamefont {Devaney}},\
  }\href@noop {} {\emph {\bibinfo {title} {{Differential Equations, Dynamical
  Systems and an introduction to Chaos}}}},\ \bibinfo {edition} {2nd}\ ed.\
  (\bibinfo  {publisher} {Elsevier},\ \bibinfo {year} {2004})\BibitemShut
  {NoStop}%
\bibitem [{\citenamefont {Bartolo}\ \emph {et~al.}(2016)\citenamefont
  {Bartolo}, \citenamefont {Minganti}, \citenamefont {Casteels},\ and\
  \citenamefont {Ciuti}}]{Bartolo2016}%
  \BibitemOpen
  \bibfield  {author} {\bibinfo {author} {\bibfnamefont {N.}~\bibnamefont
  {Bartolo}}, \bibinfo {author} {\bibfnamefont {F.}~\bibnamefont {Minganti}},
  \bibinfo {author} {\bibfnamefont {W.}~\bibnamefont {Casteels}}, \ and\
  \bibinfo {author} {\bibfnamefont {C.}~\bibnamefont {Ciuti}},\ }\href
  {http://arxiv.org/abs/1607.06739} {\  (\bibinfo {year} {2016})},\ \Eprint
  {http://arxiv.org/abs/1607.06739} {arXiv:1607.06739} \BibitemShut {NoStop}%
\bibitem [{\citenamefont {Zagoskin}\ \emph {et~al.}(2008)\citenamefont
  {Zagoskin}, \citenamefont {Il'ichev}, \citenamefont {McCutcheon},
  \citenamefont {Young},\ and\ \citenamefont {Nori}}]{Zagoskin2008}%
  \BibitemOpen
  \bibfield  {author} {\bibinfo {author} {\bibfnamefont {A.~M.}\ \bibnamefont
  {Zagoskin}}, \bibinfo {author} {\bibfnamefont {E.}~\bibnamefont {Il'ichev}},
  \bibinfo {author} {\bibfnamefont {M.~W.}\ \bibnamefont {McCutcheon}},
  \bibinfo {author} {\bibfnamefont {J.~F.}\ \bibnamefont {Young}}, \ and\
  \bibinfo {author} {\bibfnamefont {F.}~\bibnamefont {Nori}},\ }\href@noop {}
  {\bibfield  {journal} {\bibinfo  {journal} {Phys. Rev. Lett.}\ }\textbf
  {\bibinfo {volume} {101}},\ \bibinfo {pages} {253602} (\bibinfo {year}
  {2008})}\BibitemShut {NoStop}%
\bibitem [{\citenamefont {Didier}\ \emph {et~al.}(2014)\citenamefont {Didier},
  \citenamefont {Qassemi},\ and\ \citenamefont {Blais}}]{Didier2014}%
  \BibitemOpen
  \bibfield  {author} {\bibinfo {author} {\bibfnamefont {N.}~\bibnamefont
  {Didier}}, \bibinfo {author} {\bibfnamefont {F.}~\bibnamefont {Qassemi}}, \
  and\ \bibinfo {author} {\bibfnamefont {A.}~\bibnamefont {Blais}},\
  }\href@noop {} {\bibfield  {journal} {\bibinfo  {journal} {Phys. Rev. A}\
  }\textbf {\bibinfo {volume} {89}},\ \bibinfo {pages} {013820} (\bibinfo
  {year} {2014})}\BibitemShut {NoStop}%
\bibitem [{\citenamefont {Boissonneault}\ \emph {et~al.}(2014)\citenamefont
  {Boissonneault}, \citenamefont {Doherty}, \citenamefont {Ong}, \citenamefont
  {Bertet}, \citenamefont {Vion}, \citenamefont {Esteve},\ and\ \citenamefont
  {Blais}}]{Boissonneault2014}%
  \BibitemOpen
  \bibfield  {author} {\bibinfo {author} {\bibfnamefont {M.}~\bibnamefont
  {Boissonneault}}, \bibinfo {author} {\bibfnamefont {A.~C.}\ \bibnamefont
  {Doherty}}, \bibinfo {author} {\bibfnamefont {F.~R.}\ \bibnamefont {Ong}},
  \bibinfo {author} {\bibfnamefont {P.}~\bibnamefont {Bertet}}, \bibinfo
  {author} {\bibfnamefont {D.}~\bibnamefont {Vion}}, \bibinfo {author}
  {\bibfnamefont {D.}~\bibnamefont {Esteve}}, \ and\ \bibinfo {author}
  {\bibfnamefont {A.}~\bibnamefont {Blais}},\ }\href@noop {} {\bibfield
  {journal} {\bibinfo  {journal} {Phys. Rev. A}\ }\textbf {\bibinfo {volume}
  {89}},\ \bibinfo {pages} {022324} (\bibinfo {year} {2014})}\BibitemShut
  {NoStop}%
\bibitem [{\citenamefont {Elliott}\ and\ \citenamefont
  {Ginossar}(2015)}]{Elliott2015}%
  \BibitemOpen
  \bibfield  {author} {\bibinfo {author} {\bibfnamefont {M.}~\bibnamefont
  {Elliott}}\ and\ \bibinfo {author} {\bibfnamefont {E.}~\bibnamefont
  {Ginossar}},\ }\href {\doibase 10.1103/PhysRevA.92.013826} {\bibfield
  {journal} {\bibinfo  {journal} {Phys. Rev. A}\ }\textbf {\bibinfo {volume}
  {92}},\ \bibinfo {pages} {013826} (\bibinfo {year} {2015})}\BibitemShut
  {NoStop}%
\bibitem [{\citenamefont {Vlastakis}\ \emph {et~al.}(2013)\citenamefont
  {Vlastakis}, \citenamefont {Kirchmair}, \citenamefont {Leghtas},
  \citenamefont {Nigg}, \citenamefont {Frunzio}, \citenamefont {Girvin},
  \citenamefont {Mirrahimi}, \citenamefont {Devoret},\ and\ \citenamefont
  {Schoelkopf}}]{Vlastakis2013}%
  \BibitemOpen
  \bibfield  {author} {\bibinfo {author} {\bibfnamefont {B.}~\bibnamefont
  {Vlastakis}}, \bibinfo {author} {\bibfnamefont {G.}~\bibnamefont
  {Kirchmair}}, \bibinfo {author} {\bibfnamefont {Z.}~\bibnamefont {Leghtas}},
  \bibinfo {author} {\bibfnamefont {S.~E.}\ \bibnamefont {Nigg}}, \bibinfo
  {author} {\bibfnamefont {L.}~\bibnamefont {Frunzio}}, \bibinfo {author}
  {\bibfnamefont {S.~M.}\ \bibnamefont {Girvin}}, \bibinfo {author}
  {\bibfnamefont {M.}~\bibnamefont {Mirrahimi}}, \bibinfo {author}
  {\bibfnamefont {M.}~\bibnamefont {Devoret}}, \ and\ \bibinfo {author}
  {\bibfnamefont {R.~J.}\ \bibnamefont {Schoelkopf}},\ }\href@noop {}
  {\bibfield  {journal} {\bibinfo  {journal} {Science}\ }\textbf {\bibinfo
  {volume} {342}},\ \bibinfo {pages} {607} (\bibinfo {year}
  {2013})}\BibitemShut {NoStop}%
\bibitem [{\citenamefont {Mirrahimi}\ \emph {et~al.}(2014)\citenamefont
  {Mirrahimi}, \citenamefont {Leghtas}, \citenamefont {Albert}, \citenamefont
  {Touzard}, \citenamefont {Schoelkopf}, \citenamefont {Jiang},\ and\
  \citenamefont {Devoret}}]{Mirrahimi2013}%
  \BibitemOpen
  \bibfield  {author} {\bibinfo {author} {\bibfnamefont {M.}~\bibnamefont
  {Mirrahimi}}, \bibinfo {author} {\bibfnamefont {Z.}~\bibnamefont {Leghtas}},
  \bibinfo {author} {\bibfnamefont {V.~V.}\ \bibnamefont {Albert}}, \bibinfo
  {author} {\bibfnamefont {S.}~\bibnamefont {Touzard}}, \bibinfo {author}
  {\bibfnamefont {R.~J.}\ \bibnamefont {Schoelkopf}}, \bibinfo {author}
  {\bibfnamefont {L.}~\bibnamefont {Jiang}}, \ and\ \bibinfo {author}
  {\bibfnamefont {M.}~\bibnamefont {Devoret}},\ }\href {\doibase
  10.1088/1367-2630/16/4/045014} {\bibfield  {journal} {\bibinfo  {journal}
  {New J. Phys.}\ }\textbf {\bibinfo {volume} {16}},\ \bibinfo {pages} {045014}
  (\bibinfo {year} {2014})}\BibitemShut {NoStop}%
\bibitem [{\citenamefont {Ofek}\ \emph {et~al.}(2016)\citenamefont {Ofek},
  \citenamefont {Petrenko}, \citenamefont {Heeres}, \citenamefont {Reinhold},
  \citenamefont {Leghtas}, \citenamefont {Vlastakis}, \citenamefont {Liu},
  \citenamefont {Frunzio}, \citenamefont {Girvin}, \citenamefont {Jiang},
  \citenamefont {Mirrahimi}, \citenamefont {Devoret},\ and\ \citenamefont
  {Schoelkopf}}]{Ofek2016}%
  \BibitemOpen
  \bibfield  {author} {\bibinfo {author} {\bibfnamefont {N.}~\bibnamefont
  {Ofek}}, \bibinfo {author} {\bibfnamefont {A.}~\bibnamefont {Petrenko}},
  \bibinfo {author} {\bibfnamefont {R.}~\bibnamefont {Heeres}}, \bibinfo
  {author} {\bibfnamefont {P.}~\bibnamefont {Reinhold}}, \bibinfo {author}
  {\bibfnamefont {Z.}~\bibnamefont {Leghtas}}, \bibinfo {author} {\bibfnamefont
  {B.}~\bibnamefont {Vlastakis}}, \bibinfo {author} {\bibfnamefont
  {Y.}~\bibnamefont {Liu}}, \bibinfo {author} {\bibfnamefont {L.}~\bibnamefont
  {Frunzio}}, \bibinfo {author} {\bibfnamefont {S.~M.}\ \bibnamefont {Girvin}},
  \bibinfo {author} {\bibfnamefont {L.}~\bibnamefont {Jiang}}, \bibinfo
  {author} {\bibfnamefont {M.}~\bibnamefont {Mirrahimi}}, \bibinfo {author}
  {\bibfnamefont {M.~H.}\ \bibnamefont {Devoret}}, \ and\ \bibinfo {author}
  {\bibfnamefont {R.~J.}\ \bibnamefont {Schoelkopf}},\ }\href {\doibase
  10.1038/nature18949} {\bibfield  {journal} {\bibinfo  {journal} {Nature}\
  }\textbf {\bibinfo {volume} {536}},\ \bibinfo {pages} {441} (\bibinfo {year}
  {2016})}\BibitemShut {NoStop}%
\bibitem [{\citenamefont {Boissonneault}\ \emph {et~al.}(2009)\citenamefont
  {Boissonneault}, \citenamefont {Gambetta},\ and\ \citenamefont
  {Blais}}]{Boissonneault2009}%
  \BibitemOpen
  \bibfield  {author} {\bibinfo {author} {\bibfnamefont {M.}~\bibnamefont
  {Boissonneault}}, \bibinfo {author} {\bibfnamefont {J.~M.}\ \bibnamefont
  {Gambetta}}, \ and\ \bibinfo {author} {\bibfnamefont {A.}~\bibnamefont
  {Blais}},\ }\href@noop {} {\bibfield  {journal} {\bibinfo  {journal} {Phys.
  Rev. A}\ }\textbf {\bibinfo {volume} {79}},\ \bibinfo {pages} {013819}
  (\bibinfo {year} {2009})}\BibitemShut {NoStop}%
\bibitem [{\citenamefont {Goto}(2016{\natexlab{a}})}]{Goto2016}%
  \BibitemOpen
  \bibfield  {author} {\bibinfo {author} {\bibfnamefont {H.}~\bibnamefont
  {Goto}},\ }\href@noop {} {\bibfield  {journal} {\bibinfo  {journal} {Phys.
  Rev. A}\ }\textbf {\bibinfo {volume} {93}},\ \bibinfo {pages} {050301}
  (\bibinfo {year} {2016}{\natexlab{a}})}\BibitemShut {NoStop}%
\bibitem [{\citenamefont {Goto}(2016{\natexlab{b}})}]{Goto2016a}%
  \BibitemOpen
  \bibfield  {author} {\bibinfo {author} {\bibfnamefont {H.}~\bibnamefont
  {Goto}},\ }\href@noop {} {\bibfield  {journal} {\bibinfo  {journal} {Sci.
  Rep.}\ }\textbf {\bibinfo {volume} {6}},\ \bibinfo {pages} {21686} (\bibinfo
  {year} {2016}{\natexlab{b}})}\BibitemShut {NoStop}%
\bibitem [{\citenamefont {Wolinsky}\ and\ \citenamefont
  {Carmichael}(1988)}]{Wolinsky1988}%
  \BibitemOpen
  \bibfield  {author} {\bibinfo {author} {\bibfnamefont {M.}~\bibnamefont
  {Wolinsky}}\ and\ \bibinfo {author} {\bibfnamefont {H.~J.}\ \bibnamefont
  {Carmichael}},\ }\href {\doibase 10.1103/PhysRevLett.60.1836} {\bibfield
  {journal} {\bibinfo  {journal} {Phys. Rev. Lett.}\ }\textbf {\bibinfo
  {volume} {60}},\ \bibinfo {pages} {1836} (\bibinfo {year}
  {1988})}\BibitemShut {NoStop}%
\bibitem [{\citenamefont {Minganti}\ \emph {et~al.}(2016)\citenamefont
  {Minganti}, \citenamefont {Bartolo}, \citenamefont {Lolli}, \citenamefont
  {Casteels},\ and\ \citenamefont {Ciuti}}]{Minganti2016}%
  \BibitemOpen
  \bibfield  {author} {\bibinfo {author} {\bibfnamefont {F.}~\bibnamefont
  {Minganti}}, \bibinfo {author} {\bibfnamefont {N.}~\bibnamefont {Bartolo}},
  \bibinfo {author} {\bibfnamefont {J.}~\bibnamefont {Lolli}}, \bibinfo
  {author} {\bibfnamefont {W.}~\bibnamefont {Casteels}}, \ and\ \bibinfo
  {author} {\bibfnamefont {C.}~\bibnamefont {Ciuti}},\ }\href {\doibase
  10.1038/srep26987} {\bibfield  {journal} {\bibinfo  {journal} {Sci. Rep.}\
  }\textbf {\bibinfo {volume} {6}},\ \bibinfo {pages} {26987} (\bibinfo {year}
  {2016})}\BibitemShut {NoStop}%
\bibitem [{\citenamefont {Puri}\ and\ \citenamefont
  {Blais}(2016{\natexlab{b}})}]{Puri2016a}%
  \BibitemOpen
  \bibfield  {author} {\bibinfo {author} {\bibfnamefont {S.}~\bibnamefont
  {Puri}}\ and\ \bibinfo {author} {\bibfnamefont {A.}~\bibnamefont {Blais}},\
  }\href@noop {} {\  (\bibinfo {year} {2016}{\natexlab{b}})},\ \Eprint
  {http://arxiv.org/abs/arXiv:1605.09408v1} {arXiv:arXiv:1605.09408v1}
  \BibitemShut {NoStop}%
\bibitem [{\citenamefont {Kheruntsyan}(1999)}]{Kheruntsyan1999}%
  \BibitemOpen
  \bibfield  {author} {\bibinfo {author} {\bibfnamefont {K.~V.}\ \bibnamefont
  {Kheruntsyan}},\ }\href {\doibase 10.1088/1464-4266/1/2/005} {\bibfield
  {journal} {\bibinfo  {journal} {J. Opt. B Quantum Semiclassical Opt.}\
  }\textbf {\bibinfo {volume} {1}},\ \bibinfo {pages} {225} (\bibinfo {year}
  {1999})}\BibitemShut {NoStop}%
\end{thebibliography}%
\bibliographystyle{apsrev4-1}

\end{document}